\documentclass[10pt]{article}
\usepackage{graphicx}
\usepackage{amsmath}
\usepackage{amssymb}
\usepackage{caption2}
\begin{document}
\begin{center}
{\large {\bf \sc{  Revisit  the  $X(4274)$ as the axialvector tetraquark state
  }}} \\[2mm]
Zhi-Gang  Wang \footnote{E-mail: zgwang@aliyun.com.  }   \\
 Department of Physics, North China Electric Power University, Baoding 071003, P. R. China
\end{center}

\begin{abstract}
In this article, we construct  the $[sc]_A[\bar{s}\bar{c}]_V-[sc]_V[\bar{s}\bar{c}]_A$ type tensor  current to study  the mass and width of the $X(4274)$ with the QCD sum rules in details.    The  predicted mass $M_{X}=(4.27\pm0.09) \, \rm{GeV}$ for the $J^{PC}=1^{++}$ tetraquark state is in excellent agreement with the experimental data  $4273.3 \pm 8.3 ^{+17.2}_{-3.6} \mbox{ MeV}$  from the LHCb collaboration. The central value of the width $\Gamma(X(4274)\to J/\psi \phi)=47.9\,{\rm{MeV}}$ is in excellent agreement with  the experimental data  $56 \pm 11 ^{+8}_{-11} {\mbox{ MeV}}$ from the LHCb collaboration. The present work supports assigning the $X(4274)$ to be the $J^{PC}=1^{++}$  $[sc]_A[\bar{s}\bar{c}]_V-[sc]_V[\bar{s}\bar{c}]_A$   tetraquark state with a relative P-wave between the diquark and antidiquark constituents. Furthermore, we obtain the mass of the $[sc]_A[\bar{s}\bar{c}]_V-[sc]_V[\bar{s}\bar{c}]_A$ type tetraquark state with $J^{PC}=1^{-+}$ as a byproduct.
\end{abstract}

PACS number: 12.39.Mk, 12.38.Lg

Key words: Tetraquark  states, QCD sum rules

\section{Introduction}

In 2011,  the CDF collaboration confirmed the
$X(4140)$ in the $B^\pm\rightarrow J/\psi\,\phi K^\pm$ decays  produced in $p\bar{p}$  collisions at $\sqrt{s}=1.96~\rm{TeV}$ with
a  statistical significance greater  than $5\sigma$, and observed an evidence for the $X(4274)$ with approximate significance of $3.1\sigma$. The
measured mass and width
 are $4274.4^{+8.4}_{-6.7}\pm1.9\,\rm{MeV}$ and
$32.3^{+21.9}_{-15.3}\pm7.6\,\rm{MeV}$, respectively
\cite{CDF1101}. In 2013, the CMS  collaboration  observed an evidence for a second peaking structure (which is consistent with the $X(4274)$) besides the $X(4140)$ with the mass
$ 4313.8 \pm 5.3 \pm 7.3 \, \rm{MeV}$ and width $ 38^{+30}_{-15}\pm  16 \,\rm{MeV}$ respectively in the $B^\pm\rightarrow J/\psi\,\phi K^\pm$ decays produced in $pp$ collisions at $\sqrt{s}=7 \,\rm{TeV}$ collected with the CMS detector at the LHC \cite{CMS1309}.

In 2016, the LHCb collaboration performed the first full amplitude analysis of the decays $B^+\to J/\psi \phi K^+$ with a data sample of $3\rm{fb}^{-1}$ of $pp$ collision data collected at $\sqrt{s} = 7$ and $8\, \rm{ TeV}$ with the LHCb detector, and confirmed the two old particles $X(4140)$ and $X(4274)$ in the $J/\psi \phi$   mass spectrum  with statistical significances $8.4\sigma$ and $6.0\sigma$, respectively,
the  measured masses and widths are
\begin{flalign}
 & X(4140) : M_X = 4146.5 \pm 4.5 ^{+4.6}_{-2.8} \mbox{ MeV}\, , \, \Gamma_X = 83 \pm 21 ^{+21}_{-14} \mbox{ MeV} \, , \nonumber\\
 & X(4274) : M_X = 4273.3 \pm 8.3 ^{+17.2}_{-3.6} \mbox{ MeV}\, , \, \Gamma_X = 56 \pm 11 ^{+8}_{-11} \mbox{ MeV} \,  .
\end{flalign}
Furthermore, the LHCb  collaboration   determined the spin-parity-charge-conjugation  of the $X(4140)$ and $X(4274)$ to be $J^{PC} =1^{++}$ with statistical significances $5.7\sigma$ and $5.8\sigma$, respectively for the first time \cite{LHCb-16061,LHCb-16062}, which rules  out the $0^{-+}$ molecule assignment, and it is consistent with our previous work \cite{Wang4274-1102}.

There have been several possible assignments, such as the color sextet-sextet type $cs\bar{c}\bar{s}$ tetraquark state \cite{Zhu-X4140,Zhu-X4274,Azizi-X4274}, the conventional orbitally excited state $\chi_{c1}(\rm 3P)$ \cite{LiuXH-4274,Wang-2016-Y4274}, the color triplet-triplet type $ \frac{1}{\sqrt{6}}(u\bar{u}+d\bar{d}-2s\bar{s})c\bar{c}$ tetraquark state \cite{ZhuR-Y4274}, etc.
In Ref.\cite{Maiani-X4140}, L. Maiani, A. D. Polosa and V. Riquer take the mass of the $X(4140)$ as input parameter, and obtain the mass spectrum of the $sc\bar{s}\bar{c}$ tetraquark states with positive parity based on the effective  Hamiltonian with the spin-spin and spin-orbit  interactions, however, they observe that there is no room for the $X(4274)$, and suggest that  the $X(4274)$
 corresponds to two, almost degenerate, unresolved lines with $J^{PC}=0^{++}$ and $2^{++}$.

In Ref.\cite{Wang-2016-Y4274}, we  construct the  color octet-octet type axialvector current to study the mass and width of the  $X(4274)$  with the QCD sum rules in details,   the predicted mass favors assigning  the   $X(4274)$  to be  the color octet-octet  type tetraquark molecule-like  state, but the predicted width disfavors assigning  the  $X(4274)$  to be  the color octet-octet  type tetraquark molecule-like  state strongly.

In Ref.\cite{Wang-2016-Y4140},  we study the masses of the $[sc]_S[\bar{s}\bar{c}]_A\pm [sc]_A[\bar{s}\bar{c}]_S$ type and  $[sc]_P[\bar{s}\bar{c}]_V \mp [sc]_V[\bar{s}\bar{c}]_P$ type    tetraquark states with  $J^{PC}=1^{+\pm}$ respectively with the QCD sum rules in details, where the subscripts  $S$, $P$, $A$ and $V$ denote the scalar, pseudoscalar, axialvector and vector diquark constituents  respectively,  the numerical results $M_{X}=3.95\pm0.09\,\rm{GeV}$ and $5.00\pm0.10\,\rm{GeV}$ disfavor assigning the $X(4140/4274)$ to be the $J^{PC}=1^{++}$ $[sc]_S[\bar{s}\bar{c}]_A+ [sc]_A[\bar{s}\bar{c}]_S$ type and $[sc]_P[\bar{s}\bar{c}]_V-[sc]_V[\bar{s}\bar{c}]_P$ type  tetraquark states.

In Ref.\cite{WangDi-1811}, we construct  both the $[sc]_T[\bar{s}\bar{c}]_A+[sc]_A[\bar{s}\bar{c}]_T$ type and $[sc]_T[\bar{s}\bar{c}]_V-[sc]_V[\bar{s}\bar{c}]_T$ type axialvector currents with $J^{PC}=1^{++}$  to study the mass and width of the $X(4140)$ with the QCD sum rules in details, where the subscript $T$ denotes the tensor diquark operator. The predicted masses support assigning the $X(4140)$ to be the $[sc]_T[\bar{s}\bar{c}]_V-[sc]_V[\bar{s}\bar{c}]_T$ type axialvector tetraquark state, the predicted decay width $\Gamma(X(4140)\to J/\psi \phi)=86.9\pm22.6\,{\rm{MeV}}$ is in excellent agreement with the experimental data $83\pm 21^{+21}_{-14} {\mbox{ MeV}}$ from the LHCb collaboration, which also supports  assigning the $X(4140)$ to be the $[sc]_T[\bar{s}\bar{c}]_V-[sc]_V[\bar{s}\bar{c}]_T$ type axialvector tetraquark state.

In Refs.\cite{Azizi-X4274,ChenZhu-2011}, the $[sc]_S[\bar{s}\bar{c}]_A+[sc]_A[\bar{s}\bar{c}]_S$ type   and $[sc]_S^6[\bar{s}\bar{c}]^{\bar{6}}_A+[sc]_A^6[\bar{s}\bar{c}]^{\bar{6}}_S$ type tetraquark states with $J^{PC}=1^{++}$ are studied with the QCD sum rules, the criteria for choosing the
Borel windows are different from the our previous works \cite{Wang-2016-Y4274,Wang-2016-Y4140,WangDi-1811}, one can consult Sec.2 for the technical  details. The quark structures, predicted masses and widths are all shown explicitly in Table 1.

\begin{table}
\begin{center}
\begin{tabular}{|c|c|c|c|c|c|c|c|}\hline\hline
$|S^P_{sc}, S^P_{\bar{s}\bar{c}};  L; J\rangle$   &Structures                                           &$M$(GeV)      &$\Gamma$(MeV)  &References   \\ \hline

$|0^+, 1^+; 0; 1\rangle+|1^+, 0^+;  0; 1\rangle$  &$[sc]_S[\bar{s}\bar{c}]_A+[sc]_A[\bar{s}\bar{c}]_S$  &$3.95\pm0.09$ &         &\cite{Wang-2016-Y4140}  \\
$|0^-, 1^-; 0; 1\rangle-|1^-, 0^-;  0; 1\rangle$  &$[sc]_P[\bar{s}\bar{c}]_V-[sc]_V[\bar{s}\bar{c}]_P$  &$5.00\pm0.10$ &         &\cite{Wang-2016-Y4140}  \\ \hline

$|1^+, 1^+; 0; 1\rangle+|1^+, 1^+;  0; 1\rangle$  &$[sc]_T[\bar{s}\bar{c}]_A+[sc]_A[\bar{s}\bar{c}]_T$  &$5.20\pm0.11$ &         &\cite{WangDi-1811}  \\
$|1^-, 1^-; 0; 1\rangle-|1^-, 1^-;  0; 1\rangle$  &$[sc]_T[\bar{s}\bar{c}]_V-[sc]_V[\bar{s}\bar{c}]_T$  &$4.14\pm0.10$ &$86.9\pm22.6$  &\cite{WangDi-1811} \\ \hline

$|0^+, 1^+; 0; 1\rangle+|1^+, 0^+;  0; 1\rangle$  &$[sc]_S[\bar{s}\bar{c}]_A+[sc]_A[\bar{s}\bar{c}]_S$  &$4.07\pm0.10$ &         &\cite{ChenZhu-2011}  \\
$|0^+, 1^+; 0; 1\rangle+|1^+, 0^+;  0; 1\rangle$  &$[sc]_S^6[\bar{s}\bar{c}]^{\bar{6}}_A+[sc]_A^6[\bar{s}\bar{c}]^{\bar{6}}_S$  &$4.22\pm0.10$ &   &\cite{ChenZhu-2011}  \\ \hline

$|0^+, 1^+; 0; 1\rangle+|1^+, 0^+;  0; 1\rangle$  &$[sc]_S[\bar{s}\bar{c}]_A+[sc]_A[\bar{s}\bar{c}]_S$  &$4.18\pm0.12$ &$80\pm29$&\cite{Azizi-X4274}  \\
$|0^+, 1^+; 0; 1\rangle+|1^+, 0^+;  0; 1\rangle$  &$[sc]_S^6[\bar{s}\bar{c}]^{\bar{6}}_A+[sc]_A^6[\bar{s}\bar{c}]^{\bar{6}}_S$  &$4.26\pm0.12$ &$272\pm81$&\cite{Azizi-X4274}  \\ \hline

 &$[\bar{s}c]^8_P[\bar{c}s]_V^8-[\bar{s}c]^8_V[\bar{c}s]_P^8$  &$4.27\pm0.09$ &$1800$   &\cite{Wang-2016-Y4274}  \\ \hline

$|1^+, 1^-; 1; 1\rangle-|1^-, 1^+;  1; 1\rangle$  &$[sc]_A[\bar{s}\bar{c}]_V-[sc]_V[\bar{s}\bar{c}]_A$  &               &         &This work  \\ \hline\hline

\end{tabular}
\end{center}
\caption{ The structures,  masses and widths  of the $sc\bar{s}\bar{c}$ tetraquark states with $J^{PC}=1^{++}$ from the QCD sum rules, where the superscript $P$ denotes the parity, the $S$ denotes the spin, the $L$ denotes the relative angular momentum, the $J$ denotes the total angular momentum, the superscript $8$ denotes the color octet, the superscripts $6$ and $\bar{6}$ denote the color sextet and antisextet respectively. The superscripts $3$ and $\bar{3}$ for the color triplet and antitriplet are neglected for simplicity.    }
\end{table}

In this article, we extend our previous works \cite{Wang-2016-Y4274,Wang-2016-Y4140,WangDi-1811},  construct   the $[sc]_A[\bar{s}\bar{c}]_V-[sc]_V[\bar{s}\bar{c}]_A$ type tensor  current to study the mass and decay width of the $X(4274)$ with the QCD sum rules, and explore the possible assignment of the $X(4274)$ as the diquark-antidiquark type axialvector tetraquark state once more.

The article is arranged as follows:  we derive the QCD sum rules for
the mass and width of the diquark-antidiquark type axialvector tetraquark  state  $X(4274)$   in section 2 and in section 3 respectively;  section 4 is reserved for our conclusion.

\section{The mass of the $X(4274)$ as the axialvector tetraquark  state }
In the following, we write down  the two-point correlation function $\Pi_{\mu\nu\alpha\beta}(p)$  in the QCD sum rules,
\begin{eqnarray}
\Pi_{\mu\nu\alpha\beta}(p)&=&i\int d^4x e^{ip \cdot x} \langle0|T\left\{J_{\mu\nu}(x)J_{\alpha\beta}^{\dagger}(0)\right\}|0\rangle \, ,
\end{eqnarray}
where
\begin{eqnarray}
J_{\mu\nu}(x)&=&\frac{\varepsilon^{ijk}\varepsilon^{imn}}{\sqrt{2}}\Big\{s^T_j(x)C\gamma_\mu c_k(x) \bar{s}_m(x)\gamma_5\gamma_\nu C \bar{c}^T_n(x)-s^T_j(x)C\gamma_\nu\gamma_5 c_k(x) \bar{s}_m(x)\gamma_\mu C \bar{c}^T_n(x)\Big\}\, , \nonumber\\
\end{eqnarray}
 the $i$, $j$, $k$, $m$, $n$ are color indexes, the $C$ is the charge conjugation matrix.
Under charge conjugation (parity) transform $\widehat{C}$ ($\widehat{P}$), the current $J_{\mu\nu}(x)$ has the property,
\begin{eqnarray}
\widehat{C}J_{\mu\nu}(x)\widehat{C}^{-1}&=&+ J_{\mu\nu}(x) \, , \nonumber\\
\widehat{P}J_{\mu\nu}(x)\widehat{P}^{-1}&=&- J^{\mu\nu}(\tilde{x}) \, ,
\end{eqnarray}
where $x^\mu=(t,\vec{x})$ and $\tilde{x}^\mu=(t,-\vec{x})$. The current has definite charge conjugation, and couples potentially to the tetraquark states with positive charge conjugation. The component $J_{0i}(x)$ has positive parity,  while the component  $J_{ij}(x)$ has negative parity, where the space indexes $i$, $j=1$, $2$, $3$.
The current $J_{\mu\nu}$ couples potentially to both the spin-parity-charge-conjugation $J^{PC}=1^{++}$ and $1^{-+}$ tetraquark states,
\begin{eqnarray}
  \langle 0|J_{\mu\nu}(0)|X^-(p)\rangle &=& \frac{\lambda_{X^-}}{M_{X^-}} \, \varepsilon_{\mu\nu\alpha\beta} \, \varepsilon^{\alpha}p^{\beta}\, , \nonumber\\
 \langle 0|J_{\mu\nu}(0)|X^+(p)\rangle &=&\frac{\lambda_{X^+}}{M_{X^+}} \left(\varepsilon_{\mu}p_{\nu}-\varepsilon_{\nu}p_{\mu} \right)\, ,
\end{eqnarray}
the  $\varepsilon_\mu$ are the polarization vectors of the vector and axialvector tetraquark states, the $M_{X^\pm}$ and $\lambda_{X^\pm}$ are the masses and pole residues, respectively.

The scattering amplitude for one-gluon exchange  is proportional to
\begin{eqnarray}
\left(\frac{\lambda^a}{2}\right)_{ij}\left(\frac{\lambda^a}{2}\right)_{kl}&=&-\frac{N_c+1}{4N_c}t^A_{ik}t^A_{lj}+\frac{N_c-1}{4N_c}t^S_{ik}t^S_{lj} \, ,
\end{eqnarray}
where
\begin{eqnarray}
t^A_{ik}t^A_{lj}&=&\delta_{ij}\delta_{kl}-\delta_{il}\delta_{kj}=\varepsilon_{mik}\varepsilon_{mjl}\,\nonumber\\
t^S_{ik}t^S_{lj}&=&\delta_{ij}\delta_{kl}+\delta_{il}\delta_{kj}\, ,
\end{eqnarray}
the $\lambda^a$ is the  Gell-Mann matrix,  the $i$, $j$, $k$, $m$ and $l$ are color indexes, the $N_c$ is the color number.  The negative sign in front of the antisymmetric  antitriplet $\bar{3}_c$ indicates the interaction
is attractive, which favors  formation of
the diquarks in  color antitriplet,  while the positive sign in front of the symmetric sextet $6_c$ indicates
 the interaction  is repulsive,  which  disfavors  formation of
the diquarks in  color sextet. We prefer the  diquarks in  color antitriplet $\bar{3}_c$ to the diquarks in  color sextet  $6_c$ in constructing the tetraquark current operators.

The spin-dependent hypersplitting chromomagnetic  interactions $H_{cs}$ be expressed in terms of
Pauli spin matrices $\vec{\sigma}$ and $SU_c(3)$ generators $\lambda^a$  as
\begin{eqnarray}
H_{cs} &=& - \sum_{a}^8 \sum_{i>j} \vec{\sigma}_i \cdot \vec{\sigma}_j \lambda_i^a
\lambda_j^a = 8N + \frac{1}{2}C_6({\rm tot}) -\frac{4}{3} S_{\rm tot}
(S_{\rm tot}+1)
+ C_3(Q) + \frac{8}{3}S_Q(S_Q+1) \nonumber\\
&&-C_6(Q)+ C_3(\bar Q) + \frac{8}{3}S_{\bar Q}(S_{\bar Q}+1) -C_6(\bar Q)~,
\end{eqnarray}
where the $N$ is the total number of quarks, the $Q$ and $\bar{Q}$ are the diquark and antidiquark respectively,  and the $C_3$ and $C_6$ are quadratic
Casimir operators of $SU_c(3)$ and $SU_{cs}(6)$, respectively. The chromomagnetic  interaction $H_{cs}$ favors taking the scalar diquarks or "good" diquarks in color antitriplet
as the most stable building blocks of the tetraquark states \cite{Jaffe-CS,Jaffe-PRT}, however, it cannot exclude taking the axialvector  diquarks or "bad" diquarks in color antitriplet and other
diquarks  as the  building blocks of the tetraquark states, because the dominant contributions to the tetraquark masses do not originate from the chromomagnetic  interaction $H_{cs}$. We need those "bad" diquarks besides the "good" diquarks in studying the higher tetraquark states.
The calculations based on the QCD sum rules  indicate that  the favored configurations are the scalar  and axialvector  diquark states \cite{WangDiquark,WangLDiquark}, and the heavy-light scalar and axialvector diquark states have almost  degenerate masses  \cite{WangDiquark}, the heavy-light axialvector (or "bad") diquark states are not "bad".

In fact, we can obtain the four-quark interactions $T$ from the one-gluon exchange, then perform Fierz re-arrangement both in the color and Dirac-spinor  spaces to obtain the  result,
\begin{eqnarray}
T&=&\bar{Q}\gamma_\mu \frac{\lambda^a}{2}Q\, \bar{q}\gamma^\mu \frac{\lambda^a}{2}q \nonumber\\
&=&-\frac{N_c+1}{4N_c}\Big\{- q^TC\gamma_5t^AQ\,\bar{Q}\gamma_5Ct^A\bar{q}^T+q^TC t^AQ\,\bar{Q} Ct^A\bar{q}^T-\frac{1}{2} q^TC\gamma_\mu\gamma_5t^AQ\,\bar{Q}\gamma^\mu\gamma_5Ct^A\bar{q}^T\nonumber\\
&&-\frac{1}{2} q^TC\gamma_\mu t^AQ\,\bar{Q}\gamma^\mu Ct^A\bar{q}^T\Big\}\nonumber\\
&&+\frac{N_c+1}{4N_c}\Big\{- q^TC\gamma_5t^SQ\,\bar{Q}\gamma_5Ct^S\bar{q}^T+q^TC t^SQ\,\bar{Q} Ct^S\bar{q}^T-\frac{1}{2} q^TC\gamma_\mu\gamma_5t^SQ\,\bar{Q}\gamma^\mu\gamma_5Ct^S\bar{q}^T\nonumber\\
&&-\frac{1}{2} q^TC\gamma_\mu t^SQ\,\bar{Q}\gamma^\mu Ct^S\bar{q}^T\Big\} \, .
\end{eqnarray}
We can obtain the diquark operators $q^TC\gamma_5t^AQ$, $q^TC t^AQ$, $q^TC\gamma_\mu\gamma_5t^AQ$, $q^TC\gamma_\mu t^AQ$ in the attractive channels from the QCD indeed. Although we cannot obtain
the tensor diquark operators  $q^TC\sigma_{\alpha\beta}\gamma_5t^AQ$ and $q^TC\sigma_{\alpha\beta} t^AQ$ from the one-gluon exchange, they play an important role in building the tetraquark currents. In the QCD sum rules, we can take the scalar, pseudoscalar, vector, axialvector and tensor diquark and antidiquark operators as basic constituents to  construct the tetraquark currents, then calculate the two-point   and  three-point correlation functions in full QCD (not just for the chromomagnetic  interaction) to study the masses and partial decay widths, respectively, finally we confront the predictions to the experimental data to examine the structures of the tetraquark states.

We can also construct the following currents to interpolate the axialvector tetraquark states with $J^{PC}=1^{++}$,
\begin{eqnarray}
J^1_\mu(x)&=&\frac{\varepsilon^{ijk}\varepsilon^{imn}}{\sqrt{2}}\Big[s^{Tj}(x)C\gamma_5c^k(x) \bar{s}^m(x)\gamma_\mu C \bar{c}^{Tn}(x)+s^{Tj}(x)C\gamma_\mu c^k(x)\bar{s}^m(x)\gamma_5C \bar{c}^{Tn}(x) \Big] \, ,\nonumber\\
J^2_\mu(x)&=&\frac{\varepsilon^{ijk}\varepsilon^{imn}}{\sqrt{2}}\Big[s^{Tj}(x)Cc^k(x) \bar{s}^m(x)\gamma_5\gamma_\mu C \bar{c}^{Tn}(x)- s^{Tj}(x)C\gamma_\mu \gamma_5 c^k(x)\bar{s}^m(x)C \bar{c}^{Tn}(x) \Big] \, ,\nonumber\\
J_\mu^3(x)&=&\frac{\varepsilon^{ijk}\varepsilon^{imn}}{\sqrt{2}}\Big[s^{Tj}(x)C\sigma_{\mu\nu}\gamma_5 c^k(x)\bar{s}^m(x)\gamma^\nu C \bar{c}^{Tn}(x)+s^{Tj}(x)C\gamma^\nu c^k(x)\bar{s}^m(x)\gamma_5\sigma_{\mu\nu} C \bar{c}^{Tn}(x) \Big] \, , \nonumber\\
J_\mu^4(x)&=&\frac{\varepsilon^{ijk}\varepsilon^{imn}}{\sqrt{2}}\Big[s^{Tj}(x)C\sigma_{\mu\nu} c^k(x)\bar{s}^m(x)\gamma_5\gamma^\nu C \bar{c}^{Tn}(x)-s^{Tj}(x)C\gamma^\nu \gamma_5c^k(x)\bar{s}^m(x) \sigma_{\mu\nu} C \bar{c}^{Tn}(x) \Big] \, , \nonumber\\
\end{eqnarray}
the predicted masses are not consistent with the experimental value  of the mass of the $X(4274)$ \cite{Wang-2016-Y4140,WangDi-1811}, see Table 1. In Table 1, we also present the results from the diquark-antidiquark type interpolating currents with the color sextet-sextet structure \cite{Azizi-X4274,ChenZhu-2011}.

At the hadron side, we can insert  a complete set of intermediate hadronic states with
the same quantum numbers as the current operator $J_{\mu\nu}(x)$  into the
correlation function $\Pi_{\mu\nu\alpha\beta}(p)$ to obtain the hadronic representation
\cite{SVZ79,Reinders85}. After isolating the ground state
contributions of the  lowest axialvector and vector tetraquark states, we get the following results,
\begin{eqnarray}
\Pi_{\mu\nu\alpha\beta}(p)&=&\frac{\lambda_{ X^-}^2}{M_{X^-}^2\left(M_{X^-}^2-p^2\right)}\left(p^2g_{\mu\alpha}g_{\nu\beta} -p^2g_{\mu\beta}g_{\nu\alpha} -g_{\mu\alpha}p_{\nu}p_{\beta}-g_{\nu\beta}p_{\mu}p_{\alpha}+g_{\mu\beta}p_{\nu}p_{\alpha}+g_{\nu\alpha}p_{\mu}p_{\beta}\right) \nonumber\\
&&+\frac{\lambda_{ X^+}^2}{M_{X^+}^2\left(M_{X^+}^2-p^2\right)}\left( -g_{\mu\alpha}p_{\nu}p_{\beta}-g_{\nu\beta}p_{\mu}p_{\alpha}+g_{\mu\beta}p_{\nu}p_{\alpha}+g_{\nu\alpha}p_{\mu}p_{\beta}\right) +\cdots \, \, .
\end{eqnarray}

We can rewrite the correlation function $\Pi_{\mu\nu\alpha\beta}(p)$ into the following form according to Lorentz covariance,
\begin{eqnarray}
\Pi_{\mu\nu\alpha\beta}(p)&=&\Pi_{X^-}(p^2)\left(p^2g_{\mu\alpha}g_{\nu\beta} -p^2g_{\mu\beta}g_{\nu\alpha} -g_{\mu\alpha}p_{\nu}p_{\beta}-g_{\nu\beta}p_{\mu}p_{\alpha}+g_{\mu\beta}p_{\nu}p_{\alpha}+g_{\nu\alpha}p_{\mu}p_{\beta}\right) \nonumber\\
&&+\Pi_{X^+}(p^2)\left( -g_{\mu\alpha}p_{\nu}p_{\beta}-g_{\nu\beta}p_{\mu}p_{\alpha}+g_{\mu\beta}p_{\nu}p_{\alpha}+g_{\nu\alpha}p_{\mu}p_{\beta}\right) \, .
\end{eqnarray}

We project out the components $\Pi_{X^\pm}(p^2)$  by introducing the operators $P_{X^\pm}^{\mu\nu\alpha\beta}$,
\begin{eqnarray}
\widetilde{\Pi}_{X^\pm}(p^2)&=&p^2\Pi_{X^\pm}(p^2)=P_{X^\pm}^{\mu\nu\alpha\beta}\Pi_{\mu\nu\alpha\beta}(p) \, ,
\end{eqnarray}
where
\begin{eqnarray}
P_{X^-}^{\mu\nu\alpha\beta}&=&\frac{1}{6}\left( g^{\mu\alpha}-\frac{p^\mu p^\alpha}{p^2}\right)\left( g^{\nu\beta}-\frac{p^\nu p^\beta}{p^2}\right)\, , \nonumber\\
P_{X^+}^{\mu\nu\alpha\beta}&=&\frac{1}{6}\left( g^{\mu\alpha}-\frac{p^\mu p^\alpha}{p^2}\right)\left( g^{\nu\beta}-\frac{p^\nu p^\beta}{p^2}\right)-\frac{1}{6}g^{\mu\alpha}g^{\nu\beta}\, .
\end{eqnarray}

 Now  we carry out  the operator product expansion for the correlation function $\Pi_{\mu\nu\alpha\beta}(p)$ up to the vacuum condensates of dimension 10.
  We contract the quark fields $s$ and $c$ in the correlation function
$\Pi_{\mu\nu\alpha\beta}(p)$ with Wick theorem, and obtain the result,
\begin{eqnarray}
\Pi_{\mu\nu\alpha\beta}(p)&=&-\frac{i}{2} \varepsilon^{ijk}\varepsilon^{imn}\varepsilon^{i^{\prime}j^{\prime}k^{\prime}}\varepsilon^{i^{\prime}m^{\prime}n^{\prime}}   \int d^4x e^{ip \cdot x}   \nonumber\\
&&\left\{{\rm Tr}\left[ \gamma_{\mu} S_c^{kk^{\prime}}(x)\gamma_\alpha  CS^{Tjj^{\prime}}(x)C\right] {\rm Tr}\left[ \gamma_\beta\gamma_5 S_c^{n^{\prime}n}(-x)\gamma_5\gamma_\nu CS^{Tm^{\prime}m}(-x)C\right] \right. \nonumber\\
&&+{\rm Tr}\left[ \gamma_\nu\gamma_5 S_c^{kk^{\prime}}(x)\gamma_\alpha CS^{Tjj^{\prime}}(x)C\right] {\rm Tr}\left[ \gamma_\beta\gamma_5 S_c^{n^{\prime}n}(-x)\gamma_{\mu}CS^{Tm^{\prime}m}(-x)C\right] \nonumber\\
&&+ {\rm Tr}\left[ \gamma_{\mu} S_c^{kk^{\prime}}(x)\gamma_5\gamma_{\beta} CS^{Tjj^{\prime}}(x)C\right] {\rm Tr}\left[ \gamma_{\alpha} S_c^{n^{\prime}n}(-x)\gamma_5 \gamma_\nu CS^{Tm^{\prime}m}(-x)C\right] \nonumber\\
 &&\left.+ {\rm Tr}\left[ \gamma_{\nu}\gamma_5 S_c^{kk^{\prime}}(x)\gamma_5\gamma_{\beta}CS^{Tjj^{\prime}}(x)C\right] {\rm Tr}\left[\gamma_{\alpha}  S_c^{n^{\prime}n}(-x)\gamma_{\mu} CS^{Tm^{\prime}m}(-x)C\right] \right\} \, ,
\end{eqnarray}
where
\begin{eqnarray}\label{Propagator-L}
S^{ij}(x)&=& \frac{i\delta_{ij}\!\not\!{x}}{ 2\pi^2x^4}
-\frac{\delta_{ij}m_s}{4\pi^2x^2}-\frac{\delta_{ij}\langle
\bar{s}s\rangle}{12} +\frac{i\delta_{ij}\!\not\!{x}m_s
\langle\bar{s}s\rangle}{48}-\frac{\delta_{ij}x^2\langle \bar{s}g_s\sigma Gs\rangle}{192}+\frac{i\delta_{ij}x^2\!\not\!{x} m_s\langle \bar{s}g_s\sigma
 Gs\rangle }{1152}\nonumber\\
&& -\frac{ig_s G^{a}_{\alpha\beta}t^a_{ij}(\!\not\!{x}
\sigma^{\alpha\beta}+\sigma^{\alpha\beta} \!\not\!{x})}{32\pi^2x^2}  -\frac{\delta_{ij}x^4\langle \bar{s}s \rangle\langle g_s^2 GG\rangle}{27648}-\frac{1}{8}\langle\bar{s}_j\sigma^{\mu\nu}s_i \rangle \sigma_{\mu\nu}    +\cdots \, ,
\end{eqnarray}
\begin{eqnarray}\label{Propagator-H}
S_c^{ij}(x)&=&\frac{i}{(2\pi)^4}\int d^4k e^{-ik \cdot x} \left\{
\frac{\delta_{ij}}{\!\not\!{k}-m_c}
-\frac{g_sG^n_{\alpha\beta}t^n_{ij}}{4}\frac{\sigma^{\alpha\beta}(\!\not\!{k}+m_c)+(\!\not\!{k}+m_c)
\sigma^{\alpha\beta}}{(k^2-m_c^2)^2}\right.\nonumber\\
&&\left. -\frac{g_s^2 (t^at^b)_{ij} G^a_{\alpha\beta}G^b_{\mu\nu}(f^{\alpha\beta\mu\nu}+f^{\alpha\mu\beta\nu}+f^{\alpha\mu\nu\beta}) }{4(k^2-m_c^2)^5}+\cdots\right\} \, , \end{eqnarray}
\begin{eqnarray}
f^{\lambda\alpha\beta}&=&(\!\not\!{k}+m_c)\gamma^\lambda(\!\not\!{k}+m_c)\gamma^\alpha(\!\not\!{k}+m_c)\gamma^\beta(\!\not\!{k}+m_c)\, ,\nonumber\\
f^{\alpha\beta\mu\nu}&=&(\!\not\!{k}+m_c)\gamma^\alpha(\!\not\!{k}+m_c)\gamma^\beta(\!\not\!{k}+m_c)\gamma^\mu(\!\not\!{k}+m_c)\gamma^\nu(\!\not\!{k}+m_c)\, ,
\end{eqnarray}
and  $t^n=\frac{\lambda^n}{2}$ \cite{Reinders85}, then we project out the components
\begin{eqnarray}
\widetilde{\Pi}_{X^\pm}(p^2)&=&P_{X^\pm}^{\mu\nu\alpha\beta}\Pi_{\mu\nu\alpha\beta}(p) \, ,
\end{eqnarray}
and compute  the integrals both in the coordinate space and momentum space,  and obtain the correlation function at the QCD side therefore the QCD spectral densities through the  dispersion relation,
\begin{eqnarray}
\rho_{\pm}(s)&=&\frac{{\rm Im}\widetilde{\Pi}_{X^\pm}(s)}{\pi}\, .
\end{eqnarray}
For the technical details, one can consult Ref.\cite{WangHuang3900}.

Now we take the quark-hadron duality below the continuum thresholds  $s_0$ and perform Borel transform  with respect to
the variable $P^2=-p^2$ to obtain  two QCD sum rules:
\begin{eqnarray}\label{QCDSR-Res}
\lambda^2_{X^\pm}M^2_{X^\pm}\, \exp\left(-\frac{M^2_{X^\pm}}{T^2}\right)= \int_{4m_c^2}^{s_0} ds\, \rho_{\pm}(s) \, \exp\left(-\frac{s}{T^2}\right) \, ,
\end{eqnarray}
where
\begin{eqnarray}
\rho_\pm(s)&=&\rho_{0}^{\pm}(s)+\rho_{3}^{\pm}(s) +\rho_{4}^{\pm}(s)+\rho_{5}^{\pm}(s)+\rho_{6}^{\pm}(s)+\rho_{7}^{\pm}(s) +\rho_{8}^{\pm}(s)+\rho_{10}^{\pm}(s)\, ,
\end{eqnarray}

\begin{eqnarray}
\rho_0^{+}(s)&=&\frac{1}{1536\pi^6}\int dydz\, yz\left(1-y-z\right)^2 \overline{m}_c^2\left(s-\overline{m}_c^2\right)^3 \nonumber\\
&&+\frac{1}{6144\pi^6}\int dydz\,yz\left(1-y-z\right)^3 \left(s-\overline{m}_c^2\right)^2\left(33s^2-18s\overline{m}_c^2+\overline{m}_c^4\right) \, ,
\end{eqnarray}

\begin{eqnarray}
\rho_3^{+}(s)&=&-\frac{m_s\langle\bar{s}s\rangle }{96\pi^4}\int dydz\, yz \left(s-\overline{m}_c^2\right)\left(s-2\overline{m}_c^2\right) \nonumber\\
&&+\frac{m_s\langle\bar{s}s\rangle}{192\pi^4} \int dydz\, yz\left(1-y-z\right)\left(35s^2-30s\overline{m}_c^2+3\overline{m}_c^4\right) \nonumber\\
&&-\frac{7m_s m_c^2\langle\bar{s}s\rangle}{48\pi^4} \int dydz\,\left(s-\overline{m}_c^2\right)  \, ,
\end{eqnarray}

\begin{eqnarray}
\rho_{4}^{+}(s)&=&+\frac{m_c^2}{2304\pi^4}\langle\frac{\alpha_{s}GG}{\pi}\rangle \int dydz\,\frac{z\left(1-y-z\right)^2}{y^2}\left(3s-4\overline{m}_c^2\right)   \nonumber\\
&&-\frac{m_c^2}{2304\pi^4}\langle\frac{\alpha_{s}GG}{\pi}\rangle \int dydz\, \frac{z\left(1-y-z\right)^3}{y^2}
\left[5s-\overline{m}_c^2+\frac{4}{3}s^2\delta\left(s-\overline{m}_c^2\right)\right]   \nonumber\\
&&+\frac{1}{2304\pi^4}\langle\frac{\alpha_{s}GG}{\pi}\rangle \int dydz\, z\left(1-y-z\right) \left(s-\overline{m}_c^2\right)\left(4s-5\overline{m}_c^2\right) \nonumber\\
&&+\frac{1}{9216\pi^4}\langle\frac{\alpha_{s}GG}{\pi}\rangle \int dydz\, \left( 2y-y^2-8zy -5z^2+6z-1\right)\left(s-\overline{m}_c^2\right)\left(2s-\overline{m}_c^2\right) \nonumber\\
&&+\frac{1}{110592\pi^4}\langle\frac{\alpha_{s}GG}{\pi}\rangle \int dydz\, \left(1-y-z\right)\left( 2y-y^2 - 26zy - 19z^2 + 20z - 1\right)\nonumber\\
&&\left(35s^2-30s\overline{m}_c^2+3\overline{m}_c^4\right)\, ,
\end{eqnarray}

\begin{eqnarray}
\rho_5^{+}(s)&=&\frac{m_s\langle\bar{s}g_{s}\sigma Gs\rangle}{576\pi^4} \int dy \, y\left(1-y\right) \left(3s-4\widetilde{m}_c^2\right) \nonumber\\
&&-\frac{m_s\langle\bar{s}g_{s}\sigma Gs\rangle}{192\pi^4} \int dydz\, yz\left[5s-\overline{m}_c^2+\frac{4}{3}s^2\delta\left(s-\overline{m}_c^2\right)\right] \nonumber\\
&&+\frac{7m_s m_c^2\langle\bar{s}g_{s}\sigma Gs\rangle}{192\pi^4} \int dy  -\frac{m_s m_c^2 \langle\bar{s}g_{s}\sigma Gs\rangle}{384\pi^4}  \int dydz\, \frac{1}{y}  \, ,
\end{eqnarray}

\begin{eqnarray}
\rho_6^{+}(s)&=&\frac{7m_c^2\langle\bar{s}s\rangle^2}{72\pi^2} \int dy  \, ,
\end{eqnarray}

\begin{eqnarray}
\rho_{7}^{+}(s)&=&\frac{m_s m_c^2\langle\bar{s}s\rangle}{864\pi^2}\langle\frac{\alpha_{s}GG}{\pi}\rangle \int dydz\,
\frac{z}{y^2}\left(5-\frac{s}{T^2}\right)\delta\left(s-\overline{m}_c^2\right)   \nonumber\\
&&+\frac{m_s m_c^2\langle\bar{s}s\rangle}{216\pi^2}\langle\frac{\alpha_{s}GG}{\pi}\rangle \int dydz \, \frac{z\left(1-y-z\right)}{y^2}
\left(\frac{1}{4}+\frac{s}{T^2}-\frac{s^2}{T^4}\right)\delta\left(s-\overline{m}_c^2\right)   \nonumber\\
&&+\frac{7m_s m_c^2\langle\bar{s}s\rangle}{432\pi^2}\langle\frac{\alpha_{s}GG}{\pi}\rangle \int dydz\,
\frac{1}{y^2}\left(y-2+y\frac{ s}{T^2}\right)\delta\left(s-\overline{m}_c^2\right)   \nonumber\\
&&+\frac{m_s\langle\bar{s}s\rangle}{3456\pi^2} \langle\frac{\alpha_{s}GG}{\pi}\rangle \int dy\,\left[5-\frac{s}{2}\delta\left(s-\widetilde{m}_c^2\right) \right]  \nonumber\\
&&+\frac{m_s\langle\bar{s}s\rangle}{3456\pi^2}\langle\frac{\alpha_{s}GG}{\pi}\rangle \int dy\,  \left[1+\frac{s}{2}\delta\left(s-\widetilde{m}_c^2\right)\right] \nonumber\\
&&-\frac{m_s\langle\bar{s}s\rangle}{1728\pi^2}\langle\frac{\alpha_{s}GG}{\pi}\rangle \int dydz\,\left[1+\frac{s}{2}\delta\left(s-\overline{m}_c^2\right)\right] \nonumber\\
&&+\frac{m_s\langle\bar{s}s\rangle}{2304\pi^2}\langle\frac{\alpha_{s}GG}{\pi}\rangle \int dydz\,
\left(y + 4 z - 1\right)\left[1+\frac{4}{3}\left(s+\frac{s^2}{T^2}\right)\delta\left(s-\overline{m}_c^2\right)\right] \nonumber\\
&&+\frac{m_s m_c^2\langle\bar{s}s\rangle}{576\pi^2}\langle\frac{\alpha_{s}GG}{\pi}\rangle \int dydz\,
\frac{1}{yz} \delta\left(s-\overline{m}_c^2\right) \nonumber\\
&&-\frac{7m_s m_c^2 \langle\bar{s}s\rangle}{1728\pi^2} \langle\frac{\alpha_{s}GG}{\pi}\rangle  \int dy\,
\left(1+\frac{s}{T^2}\right)\delta\left(s-\widetilde{m}_c^2\right) \, ,
\end{eqnarray}

\begin{eqnarray}
\rho_8^{+}(s)&=&-\frac{7m_c^2\langle\bar{s}s\rangle\langle\bar{s}g_{s}\sigma Gs\rangle}{144\pi^2}\int dy\, \left(1+\frac{s}{T^2}\right)\delta\left(s-\widetilde{m}_c^2\right) \nonumber\\
&&+\frac{m_c^2\langle\bar{s}s\rangle \langle\bar{s}g_{s}\sigma Gs\rangle}{288\pi^2} \int dy\,\frac{1}{y}\delta\left(s-\widetilde{m}_c^2\right) \, ,
\end{eqnarray}

\begin{eqnarray}
\rho_{10}^{+}(s)&=&\frac{7m_c^2\langle\bar{s}g_{s}\sigma Gs\rangle^2}{1152\pi^2T^6} \int dy\, s^2\delta\left(s-\widetilde{m}_c^2\right)\nonumber\\
&&-\frac{7 m_c^2\langle\bar{s}s\rangle^2}{324T^2}\langle\frac{\alpha_{s}GG}{\pi}\rangle \int dy\,
\frac{1}{y^2}\left(1-y\frac{ s}{2T^2}\right)\delta\left(s-\widetilde{m}_c^2\right)\nonumber\\
&&-\frac{m_c^2\langle\bar{s}s\rangle^2}{864T^2}\langle\frac{\alpha_{s}GG}{\pi}\rangle \int dy\,\frac{1}{y\left(1-y\right)} \delta\left(s-\widetilde{m}_c^2\right) \nonumber\\
&&-\frac{m_c^2 \langle\bar{s}g_{s}\sigma Gs\rangle^2}{1152\pi^2T^4} \int dy\, \frac{1}{y}s\delta\left(s-\widetilde{m}_c^2\right)\nonumber\\
&&-\frac{11m_c^2\langle\bar{s}g_{s}\sigma Gs\rangle^2}{9216\pi^2T^2} \int dy\, \frac{1}{y\left(1-y\right)} \delta\left(s-\widetilde{m}_c^2\right)\nonumber\\
&&+\frac{7m_c^2 \langle\bar{s}s\rangle^2}{1296T^6} \langle\frac{\alpha_{s}GG}{\pi}\rangle \int dy\, s^2\delta\left(s-\widetilde{m}_c^2\right)\, ,
\end{eqnarray}

\begin{eqnarray}
\rho_{0}^-(s)&=&\frac{1}{1536\pi^6}\int dydz\, yz\left(1-y-z\right)^2\left(s-\overline{m}_c^2\right)^3\left(6s-\overline{m}_c^2\right) \nonumber\\
&&+\frac{1}{6144\pi^6}\int dydz\,yz\left(1-y-z\right)^3 \left(s-\overline{m}_c^2\right)^2\left(33s^2-18s\overline{m}_c^2+\overline{m}_c^4\right) \, ,
\end{eqnarray}

\begin{eqnarray}
\rho_{3}^-(s)&=&\frac{m_s\langle\bar{s}s\rangle}{96\pi^4} \int dydz\, yz\left(s-\overline{m}_c^2\right)\left(7s-2\overline{m}_c^2\right) \nonumber\\
&&+\frac{m_s\langle\bar{s}s\rangle}{192\pi^4} \int dydz\, yz\left(1-y-z\right)\left(35s^2-30s\overline{m}_c^2+3\overline{m}_c^4\right) \nonumber\\
&&+\frac{3m_s m_c^2\langle\bar{s}s\rangle}{16\pi^4} \int dydz\, \left(s-\overline{m}_c^2\right) \, ,
\end{eqnarray}

\begin{eqnarray}
\rho_{4}^-(s)&=&-\frac{m_c^2}{2304\pi^4}\langle\frac{\alpha_{s}GG}{\pi}\rangle \int dydz\, \frac{z\left(1-y-z\right)^2}{y^2}\left(9s-4\overline{m}_c^2\right)  \nonumber\\
&&-\frac{m_c^2}{2304\pi^4}\langle\frac{\alpha_{s}GG}{\pi}\rangle \int dydz\, \frac{z\left(1-y-z\right)^3}{y^2}
\left[5s-\overline{m}_c^2+\frac{4}{3}s^2\delta\left(s-\overline{m}_c^2\right)\right]   \nonumber\\
&&-\frac{1}{2304\pi^4}\langle\frac{\alpha_{s}GG}{\pi}\rangle \int dydz\, z\left(1-y-z\right)\left(s-\overline{m}_c^2\right)\left(20s-7\overline{m}_c^2\right) \nonumber\\
&&+\frac{1}{4608\pi^4}\langle\frac{\alpha_{s}GG}{\pi}\rangle \int dydz\,\left(y^2 + (8z-2)y + 5z^2 - 6z + 1\right) \left(s-\overline{m}_c^2\right)\left(2s-\overline{m}_c^2\right) \nonumber\\
&&+\frac{1}{110592\pi^4}\langle\frac{\alpha_{s}GG}{\pi}\rangle \int dydz\, \left(1-y-z\right)\left( 2y-y^2 - 26zy - 19z^2 + 20z - 1\right)\nonumber\\
&&\left(35s^2-30s\overline{m}_c^2+3\overline{m}_c^4\right)\, ,
\end{eqnarray}

\begin{eqnarray}
\rho_{5}^-(s)&=&-\frac{m_s\langle\bar{s}g_{s}\sigma Gs\rangle}{576\pi^4} \int dy \, y\left(1-y\right) \left(9s-4\widetilde{m}_c^2\right)\nonumber\\
&&-\frac{m_s\langle\bar{s}g_{s}\sigma Gs\rangle}{192\pi^4} \int dydz\, yz\left[5s-\overline{m}_c^2+\frac{4}{3}s^2\delta\left(s-\overline{m}_c^2\right)\right] \nonumber\\
&&-\frac{3m_s m_c^2\langle\bar{s}g_{s}\sigma Gs\rangle}{64\pi^4} \int dy +\frac{m_s m_c^2 \langle\bar{s}g_{s}\sigma Gs\rangle}{384\pi^4}  \int dydz\, \frac{1}{y} \, ,
\end{eqnarray}

\begin{eqnarray}
\rho_{6}^-(s)&=&-\frac{m_c^2\langle\bar{s}s\rangle^2}{8\pi^2} \int dy  \, ,
\end{eqnarray}

\begin{eqnarray}
\rho_{7}^-(s)&=&+\frac{m_s m_c^2\langle\bar{s}s\rangle}{864\pi^2}\langle\frac{\alpha_{s}GG}{\pi}\rangle \int dydz\, \frac{z}{y^2}
\left(1-\frac{5s}{T^2}\right)\delta\left(s-\overline{m}_c^2\right)   \nonumber\\
&&+\frac{m_s m_c^2\langle\bar{s}s\rangle}{216\pi^2}\langle\frac{\alpha_{s}GG}{\pi}\rangle \int dydz \, \frac{z\left(1-y-z\right)}{y^2}
\left(\frac{1}{4}+\frac{s}{T^2}-\frac{s^2}{T^4}\right)\delta\left(s-\overline{m}_c^2\right)   \nonumber\\
&&+\frac{m_s m_c^2\langle\bar{s}s\rangle}{48\pi^2}\langle\frac{\alpha_{s}GG}{\pi}\rangle \int dydz\,
\frac{1}{y^2}\left(2-y-y\frac{ s}{T^2}\right)\delta\left(s-\overline{m}_c^2\right)   \nonumber\\
&&-\frac{m_s\langle\bar{s}s\rangle}{3456\pi^2} \langle\frac{\alpha_{s}GG}{\pi}\rangle \int dy\,\left[7+\frac{13}{2}s\delta\left(s-\widetilde{m}_c^2\right) \right]   \nonumber\\
&&-\frac{m_s\langle\bar{s}s\rangle}{1728\pi^2}\langle\frac{\alpha_{s}GG}{\pi}\rangle \int dy\,  \left[1+\frac{s}{2}\delta\left(s-\widetilde{m}_c^2\right)\right] \nonumber\\
&&+\frac{m_s\langle\bar{s}s\rangle}{864\pi^2}\langle\frac{\alpha_{s}GG}{\pi}\rangle \int dydz\,
\left[1+\frac{s}{2}\delta\left(s-\overline{m}_c^2\right)\right] \nonumber\\
&&+\frac{m_s\langle\bar{s}s\rangle}{2304\pi^2}\langle\frac{\alpha_{s}GG}{\pi}\rangle \int dydz\,
\left(y + 4 z - 1\right)\left[1+\frac{4}{3}\left(s+\frac{s^2}{T^2}\right)\delta\left(s-\overline{m}_c^2\right)\right] \nonumber\\
&&-\frac{m_s m_c^2\langle\bar{s}s\rangle}{576\pi^2}\langle\frac{\alpha_{s}GG}{\pi}\rangle \int dydz\,
\frac{1}{yz} \delta\left(s-\overline{m}_c^2\right) \nonumber\\
&&+\frac{m_s m_c^2 \langle\bar{s}s\rangle}{192\pi^2} \langle\frac{\alpha_{s}GG}{\pi}\rangle  \int dy\,\left(1+\frac{s}{T^2}\right)\delta\left(s-\widetilde{m}_c^2\right) \, ,
\end{eqnarray}

\begin{eqnarray}
\rho_{8}^-(s)&=&+\frac{m_c^2\langle\bar{s}s\rangle\langle\bar{s}g_{s}\sigma Gs\rangle}{16\pi^2} \int dy\,\left(1+\frac{s}{ T^2}\right)\delta\left(s-\widetilde{m}_c^2\right) \nonumber\\
&&-\frac{m_c^2\langle\bar{s}s\rangle \langle\bar{s}g_{s}\sigma Gs\rangle}{288\pi^2} \int dy\, \frac{1}{y}\delta\left(s-\widetilde{m}_c^2\right) \, ,
\end{eqnarray}

\begin{eqnarray}
\rho_{10}^-(s)&=&-\frac{m_c^2\langle\bar{s}g_{s}\sigma Gs\rangle^2}{128\pi^2T^6} \int dy\,s^2\delta\left(s-\widetilde{m}_c^2\right)\nonumber\\
&&+ \frac{m_c^2\langle\bar{s}s\rangle^2}{36T^2}\langle\frac{\alpha_{s}GG}{\pi}\rangle \int dy\, \frac{1}{y^2}\left(1-y\frac{ s}{2T^2}\right)\delta\left(s-\widetilde{m}_c^2\right)\nonumber\\
&&+\frac{m_c^2\langle\bar{s}s\rangle^2}{864T^2}\langle\frac{\alpha_{s}GG}{\pi}\rangle \int dy\, \frac{1}{y\left(1-y\right)}\delta\left(s-\widetilde{m}_c^2\right)\nonumber\\
&&+\frac{m_c^2 \langle\bar{s}g_{s}\sigma Gs\rangle^2}{1152\pi^2T^4} \int dy\, \frac{1}{y}s\delta\left(s-\widetilde{m}_c^2\right)\nonumber\\
&&+\frac{11m_c^2\langle\bar{s}g_{s}\sigma Gs\rangle^2}{9216\pi^2T^2} \int dy\, \frac{1}{y\left(1-y\right)}\delta\left(s-\widetilde{m}_c^2\right)\nonumber\\
&&-\frac{m_c^2}{144T^6} \langle\bar{s}s\rangle^2 \langle\frac{\alpha_{s}GG}{\pi}\rangle \int dy\, s^2\delta\left(s-\widetilde{m}_c^2\right)\, ,
\end{eqnarray}
where $\int dydz=\int_{y_i}^{y_f}dy \int_{z_i}^{1-y}dz$, $\int dy=\int_{y_i}^{y_f}dy$, $y_{f}=\frac{1+\sqrt{1-4m_c^2/s}}{2}$,
$y_{i}=\frac{1-\sqrt{1-4m_c^2/s}}{2}$, $z_{i}=\frac{y
m_c^2}{y s -m_c^2}$, $\overline{m}_c^2=\frac{(y+z)m_c^2}{yz}$,
$ \widetilde{m}_c^2=\frac{m_c^2}{y(1-y)}$, $\int_{y_i}^{y_f}dy \to \int_{0}^{1}dy$, $\int_{z_i}^{1-y}dz \to \int_{0}^{1-y}dz$,  when the $\delta$ functions $\delta\left(s-\overline{m}_c^2\right)$ and $\delta\left(s-\widetilde{m}_c^2\right)$ appear.

 We derive   Eq.\eqref{QCDSR-Res} with respect to  $\tau=\frac{1}{T^2}$, then eliminate the
 pole residues  $\lambda_{X^\pm}$ to obtain the QCD sum rules for the tetraquark masses,
 \begin{eqnarray}
M_{X^\pm}^2&=&- \frac{\int_{4m_c^2}^{s_0} ds \, \frac{d}{d \tau }\,\rho_{\pm}(s)e^{-\tau s}}{\int_{4m_c^2}^{s_0} ds \rho_{\pm}(s)e^{-\tau s}}\, .
\end{eqnarray}

At the QCD side, we take the vacuum condensates  to be the standard values
$\langle\bar{q}q \rangle=-(0.24\pm 0.01\, \rm{GeV})^3$,  $\langle\bar{s}s \rangle=(0.8\pm0.1)\langle\bar{q}q \rangle$,
 $\langle\bar{s}g_s\sigma G s \rangle=m_0^2\langle \bar{s}s \rangle$,
$m_0^2=(0.8 \pm 0.1)\,\rm{GeV}^2$, $\langle \frac{\alpha_s
GG}{\pi}\rangle=(0.33\,\rm{GeV})^4 $    at the energy scale  $\mu=1\, \rm{GeV}$
\cite{SVZ79,Reinders85,ColangeloReview}, and  take the $\overline{MS}$ masses $m_{c}(m_c)=(1.275\pm0.025)\,\rm{GeV}$ and $m_s(\mu=2\,\rm{GeV})=(0.095\pm0.005)\,\rm{GeV}$
 from the Particle Data Group \cite{PDG}.
Moreover,  we take into account
the energy-scale dependence of  the quark condensate, mixed quark condensate and $\overline{MS}$ masses according to  the renormalization group equation,
 \begin{eqnarray}
 \langle\bar{s}s \rangle(\mu)&=&\langle\bar{s}s \rangle({\rm 1 GeV})\left[\frac{\alpha_{s}({\rm 1 GeV})}{\alpha_{s}(\mu)}\right]^{\frac{12}{33-2n_f}}\, , \nonumber\\
 \langle\bar{s}g_s \sigma Gs \rangle(\mu)&=&\langle\bar{s}g_s \sigma Gs \rangle({\rm 1 GeV})\left[\frac{\alpha_{s}({\rm 1 GeV})}{\alpha_{s}(\mu)}\right]^{\frac{2}{33-2n_f}}\, ,\nonumber\\
m_c(\mu)&=&m_c(m_c)\left[\frac{\alpha_{s}(\mu)}{\alpha_{s}(m_c)}\right]^{\frac{12}{33-2n_f}} \, ,\nonumber\\
m_s(\mu)&=&m_s({\rm 2GeV} )\left[\frac{\alpha_{s}(\mu)}{\alpha_{s}({\rm 2GeV})}\right]^{\frac{12}{33-2n_f}}\, ,\nonumber\\
\alpha_s(\mu)&=&\frac{1}{b_0t}\left[1-\frac{b_1}{b_0^2}\frac{\log t}{t} +\frac{b_1^2(\log^2{t}-\log{t}-1)+b_0b_2}{b_0^4t^2}\right]\, ,
\end{eqnarray}
  where $t=\log \frac{\mu^2}{\Lambda^2}$, $b_0=\frac{33-2n_f}{12\pi}$, $b_1=\frac{153-19n_f}{24\pi^2}$, $b_2=\frac{2857-\frac{5033}{9}n_f+\frac{325}{27}n_f^2}{128\pi^3}$,  $\Lambda=210\,\rm{MeV}$, $292\,\rm{MeV}$  and  $332\,\rm{MeV}$ for the flavors  $n_f=5$, $4$ and $3$, respectively \cite{PDG,Narison-mix}, and evolve all the input parameters to the ideal  energy scales   $\mu$  to extract the masses of the tetraquark states. In this article, we choose the flavor $n_f=4$.

The hidden-charm (and hidden-bottom) tetraquark states  $qQ\bar{q}^{\prime}\bar{Q}$ can be described by a double-well potential.   In the tetraquark states $qQ\bar{q}^{\prime}\bar{Q}$,
 the $Q$-quark serves as a static well potential and  attracts  the light quark $q$  to form a heavy diquark in  color antitriplet,
 the $\bar{Q}$-quark serves  as another static well potential and attracts the light antiquark $\bar{q}^\prime$  to form a heavy antidiquark  in  color triplet \cite{Wang-tetra-formula,Wang-tetra-NPA,WangHuang-molecule}.
 The   diquark and antidiquark  attract  each other    to form a compact tetraquark state \cite{Wang-tetra-formula,Wang-tetra-NPA,WangHuang-molecule},
the two heavy quarks $Q$ and $\bar{Q}$ stabilize the tetraquark state $qQ\bar{q}^{\prime}\bar{Q}$, just as in the case
of the $(\mu^-e^+)(\mu^+ e^-)$ molecule in QED \cite{Brodsky-2014}.

We can divide the tetraquark states  $qQ\bar{q}^{\prime}\bar{Q}$ into the heavy and light degrees of freedom, the heavy degree of freedom is characterized by the effective heavy quark masses ${\mathbb{M}}_Q$,  the light degree of freedom is characterized by the virtuality $V=\sqrt{M^2_{X/Y/Z}-(2{\mathbb{M}}_Q)^2}$ which includes   the interactions among the light quarks and gluons. If there exists a P-wave between the light quark and heavy quark in the heavy diquark or between the light antiquark and heavy antiquark in the heavy antidiquark, the P-wave effect can be taken as   the  light degree of freedom, the virtuality $V=\sqrt{M^2_{X/Y/Z}-(2{\mathbb{M}}_Q)^2}$.
 On the other hand, if there exists a P-wave between the heavy diquark and heavy antidiquark,  the P-wave effect can be taken as  the  heavy degree of freedom,
  the virtuality  $V=\sqrt{M^2_{X/Y/Z}-(2{\mathbb{M}}_Q+ \rm P_E)^2}$, the energy exciting a P-wave costs about $0.5\,\rm{GeV}$, i.e. $\rm P_E\approx0.5\,\rm{GeV}$.

  In this article, we study the heavy-diquark-heavy-antidiquark type tetraquark states, in other words, the color $\bar{3}_c\otimes3_c$ type  tetraquark states, just-like the charmonium and bottomnium states, where the $\bar{Q}Q$ states are of the color $\bar{3}_c\otimes3_c$ type. For the charmonium states, the energy exciting a P-wave costs  $457\,\rm{MeV}$ \cite{PDG},
  \begin{eqnarray}
 {\rm P_E}&=& \frac{5m_{\chi_{c2}}+3m_{\chi_{c1}}+m_{\chi_{c0}}}{9}-\frac{3m_{J/\psi}+m_{\eta_c}}{4}=457\,\rm{MeV}\, .
  \end{eqnarray}
If we take the updated value ${\mathbb{M}}_c=1.82\,\rm{GeV}$ \cite{WangEPJC-1601}, then $2{\mathbb{M}}_c+ \rm P_E=4.10\,\rm{GeV}$, the energy of the heavy degree of freedom of the diquark-andidiquark type tetraquark states  $qc\bar{q}^{\prime}\bar{c}$ is estimated to be   $4.10\,\rm{GeV}$.

  We set the energy scale $\mu=V$ to obtain the ideal energy scales of the QCD spectral densities   \cite{Wang-tetra-formula,Wang-tetra-NPA,WangHuang-molecule,Wang-Vector-tetra}. The energy scale formula works well for the hidden-charm (and hidden-bottom) tetraquark states, for example, $X^*(3860)$, $X(3872)$, $Z_c(3900/3885)$, $X(3915)$, $Z_c(4020/4025)$, $X(4140)$, $Z_c(4250)$, $X(4360)$, $Z_c(4430)$,  $X(4500)$, $X(4660/4630)$,  $X(4700)$,  $Z_b(10610)$, $Z_b(10650)$, and also works well for the hidden-charm pentaquark states, for example, $P_c(4380)$ and $P_c(4450)$  \cite{Wang1508-EPJC-penta-1,Wang1508-EPJC-penta-4312}. The energy scale formula can enhance the pole contributions remarkably, and can improve the convergent behaviors of the operator product expansion.
 In 2015, we studied the scalar-diquark-scalar-diquark-antiquark type  pentaquark states with the QCD sum rules by carrying out the operator product expansion up to the vacuum condensates of dimension $10$ \cite{Wang1508-EPJC-penta-4312}. In calculations, we took  the energy scale  formula $\mu=\sqrt{M^2_{P}-(2{\mathbb{M}}_c)^2}$ to determine the ideal energy scales of the QCD spectral densities with the old value ${\mathbb{M}}_c=1.80\,\rm{GeV}$ and obtained the mass
$M_P=4.29\pm0.13\,\rm{GeV}$ for the pentaquark state with $J^P={\frac{1}{2}}^-$, which is in excellent agreement with the value of the mass
  of the new pentaquark candidate $P_c(4312)$,  $4311.9\pm0.7^{+6.8}_{-0.6} \mbox{ MeV}$,  observed by the LHCb collaboration this year \cite{LHCb-Pc4312}. Recently, we restudied the scalar-diquark-scalar-diquark-antiquark type  pentaquark states with the QCD sum rules by carrying out the operator product expansion up to the vacuum condensates of dimension $13$ and took the updated value ${\mathbb{M}}_c=1.82\,\rm{GeV}$ \cite{WangEPJC-1601}, and obtained even better pentaquark mass  $
  4.31\pm0.11\,\rm{GeV}$ \cite{WangPc-4312}.

In Ref.\cite{Wang-Vector-tetra},   we introduce the relative P-wave between the diquark and antidiquark constituents explicitly to construct the vector tetraquark currents,
 and take the modified energy scale formula  $\mu=\sqrt{M^2_{X/Y/Z}-(2{\mathbb{M}}_c+\rm P_E)^2}=\sqrt{M^2_{X/Y/Z}-(4.10\,\rm{GeV})^2}$ to determine the optimal energy scales of the QCD spectral densities, and study the vector tetraquark states  with the QCD sum rules systematically,
 the  predictions support  assigning the $Y(4220/4260)$, $Z_c(4250)$, $Y(4320/4360)$ and $Y(4390)$ to be the vector tetraquark   states.

The  axialvector diquark operator $\varepsilon^{ijk}s^T_j(x)C\gamma_\mu c_k(x)$ has the  $J^P=1^+$, while the vector  diquark operator $\varepsilon^{ijk}s^T_j(x)C\gamma_\mu\gamma_5 c_k(x)$ has the $J^P=1^-$, the tetraquark quark states $X^-$ and $X^+$ have  negative parity and positive parity respectively,
 parity conservation requires that $1^{+} +1^{-} \to 1^{-}$ for the $X^-$ and $1^{+} +1^{-} +1^{-} \to 1^{+}$ for the $X^+$, there should exist an additional P-wave (or $1^-$) between the diquark and antidiquark constituents  in the tetraquark state $X^+$.
We choose the energy scale formula
\begin{eqnarray}
\mu&=&\sqrt{M^2_{X}-(3.64\,\rm{GeV})^2}\, ,
\end{eqnarray}
 for the tetraquark state $X^-$, where we have take the updated  value ${\mathbb{M}}_c=1.82\,\rm{GeV}$ \cite{WangEPJC-1601},
\begin{eqnarray}
\mu&=&\sqrt{M^2_{X}-(4.10\,\rm{GeV})^2}\, ,
\end{eqnarray}
 for the tetraquark state $X^+$ as there exists a P-wave between the heavy diquark and heavy antidiquark constituents. If the $X(4274)$ can be assigned to be the $X^+$, the optimal energy scale of the QCD spectral density is $\mu=1.2\,\rm{GeV}$. At the energy scale $\mu=1.2\,\rm{GeV}$, the flavor $SU(3)$ breaking effects are sizeable, we take into account the effect of the finite $s$ quark mass, and take the energy scale $\mu=\sqrt{M^2_{X}-(4.10\,\rm{GeV})^2}-2m_s(\mu)\approx 1\,\rm{GeV}$ in calculations.
 We evolve   all the input parameters in the QCD spectral densities to the special energy scales determined by the energy scale formula,  as the integrals
 \begin{eqnarray}
 \int_{4m_c^2(\mu)}^{s_0} ds \rho_{\pm}(s,\mu)\exp\left(-\frac{s}{T^2} \right)\, ,
 \end{eqnarray}
are sensitive to the heavy quark mass $m_c$   or the energy scale $\mu$.
In calculations, we observe that the predicted masses decrease monotonously and quickly with increase of the energy scales.
If we
abandon the energy scale formula $\mu=\sqrt{M^2_{X/Y/Z}-(2{\mathbb{M}}_c})^2$ or modified energy scale formula $\mu=\sqrt{M^2_{X/Y/Z}-(2{\mathbb{M}}_c+{\rm P_E})^2}$, we are puzzled about which energy scale should be chosen. With the help of the (modified) energy scale formula, we can choose the acceptable or optimal energy scales of the QCD spectral densities in a consistent way, the values of the effective heavy quark masses ${\mathbb{M}}_Q$ are universal for all the diquark-antidiquark type hidden-charm  and hidden-bottom tetraquark states \cite{Wang-tetra-formula,Wang-tetra-NPA,WangEPJC-1601,Wang-Vector-tetra}.

\begin{table}
\begin{center}
\begin{tabular}{|c|c|c|c|c|c|c|c|}\hline\hline
        &$T^2(\rm{GeV}^2)$   &$\sqrt{s_0}(\rm{GeV})$  &$\mu(\rm{GeV})$  &pole          &$M(\rm{GeV})$  &$\lambda(\rm{GeV}^5)$ \\ \hline

$X^+$   &$2.9-3.3$           &$4.80\pm0.10$           &1.0              &$(38-60)\%$   &$4.27\pm0.09$  &$(1.52\pm0.25)\times10^{-2}$   \\ \hline
$X^-$   &$3.7-4.3$           &$5.15\pm0.10$           &2.9              &$(39-61)\%$   &$4.66\pm0.08$  &$(7.94\pm1.00)\times10^{-2}$   \\ \hline
\end{tabular}
\end{center}
\caption{ The Borel  windows, continuum threshold parameters, ideal energy scales, pole contributions,   masses and pole residues for the axialvector and vector
  tetraquark states. }
\end{table}

\begin{figure}
 \centering
 \includegraphics[totalheight=5cm,width=7cm]{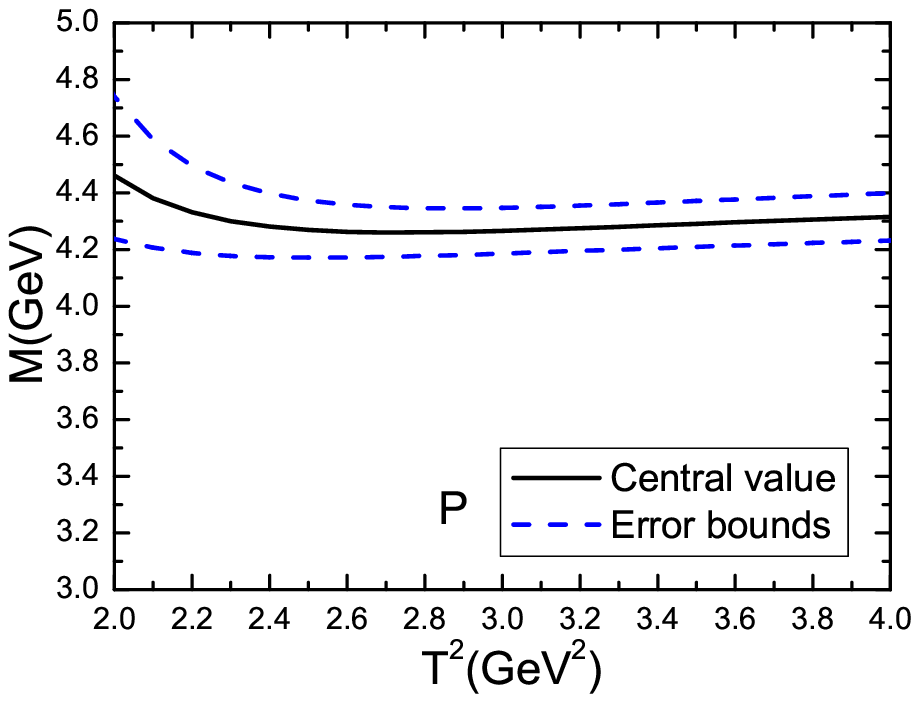}
 \includegraphics[totalheight=5cm,width=7cm]{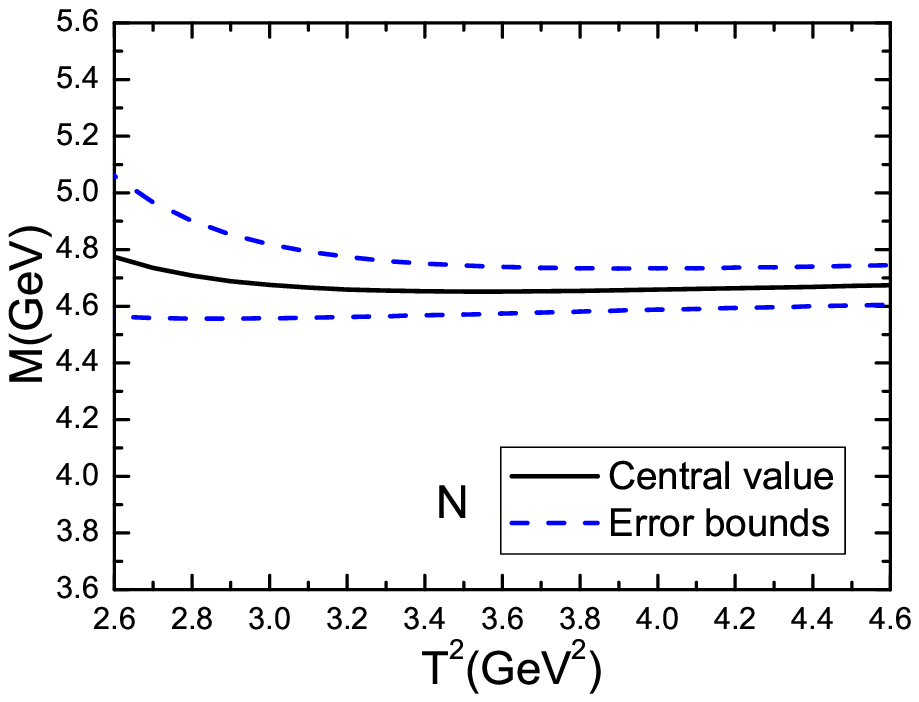}
   \caption{ The masses of the  tetraquark states with variations of the Borel parameters $T^2$, where  the P and N denote to the positive parity and negative parity, respectively.   }
\end{figure}

\begin{figure}
 \centering
 \includegraphics[totalheight=5cm,width=7cm]{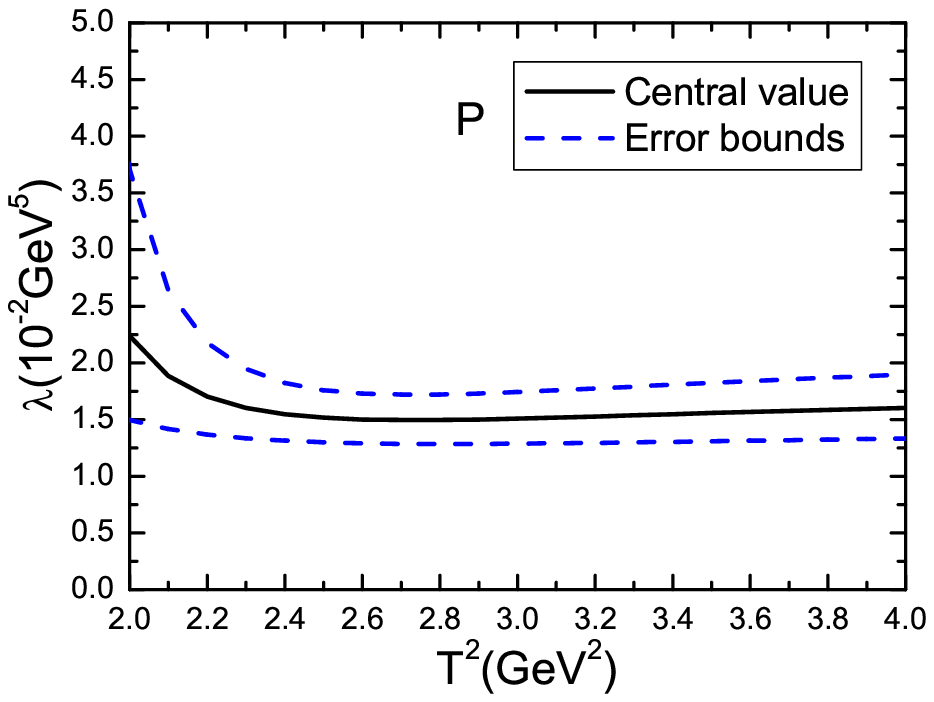}
 \includegraphics[totalheight=5cm,width=7cm]{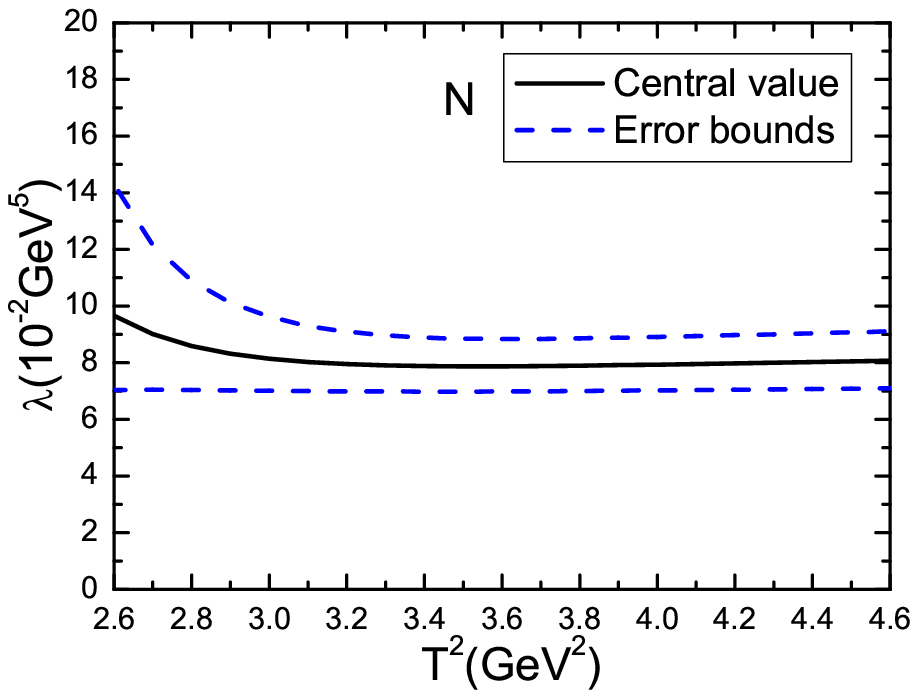}
   \caption{ The pole residues of the  tetraquark states with variations of the Borel parameters $T^2$, where  the P and N denote to the positive parity and negative parity, respectively.     }
\end{figure}

Now we search for the optimal  Borel parameters $T^2$ and continuum threshold parameters $s_0$  to satisfy   the  following four criteria:\\
$\bf 1.$ Pole dominance at the phenomenological side;\\
$\bf 2.$ Convergence of the operator product expansion;\\
$\bf 3.$ Appearance of the Borel platforms;\\
$\bf 4.$ Satisfying the  energy scale formula,\\
  via try and error, and  obtain the Borel windows $T^2$, continuum threshold parameters $s_0$, optimal  energy scales of the QCD spectral densities, and pole contributions of the ground states, which are shown   explicitly in Table 2.

From Table 2, we can see that the pole contributions are about $(40-60)\%$, the pole dominance criterion is well satisfied. In calculations, we observe that
the contributions of the vacuum condensates of  dimension 10  are about $1\%$ and $\ll1\%$ for the tetraquark states $X^+$ and $X^-$, respectively, the operator product expansion is well convergent. The first two criteria or the basic criteria of the QCD sum rules are satisfied.

We take into account all uncertainties of the input parameters, and obtain the values of the  masses and pole residues of the $sc\bar{s}\bar{c}$ tetraquark states, which are shown explicitly in Figs.1-2 and Table 2,
 \begin{eqnarray}\label{mass}
 M_{X^+}&=&(4.27\pm0.09) \, \rm{GeV}\, , \nonumber\\
 M_{X^-}&=&(4.66\pm0.08) \, \rm{GeV}\, ,  \\
 \lambda_{X^+}&=&(1.52\pm0.25)\times 10^{-2} \, \rm{GeV}^5\, , \nonumber\\
 \lambda_{X^-}&=&(7.94\pm1.00)\times 10^{-2} \, \rm{GeV}^5\, .
 \end{eqnarray}

In Figs.1-2,  we plot the masses and pole residues of the tetraquark states   with variations of the  Borel parameters $T^2$ at  larger intervals than the Borel windows. From the figure, we can see that there appear platforms in the Borel windows, the criterion $\bf{3}$  is also satisfied. From Table 2, we can see that the energy scale formula is satisfied. Now the four  criteria are all satisfied, it is reliable to extract the ground state masses. The predicted mass $m_{X^+}=(4.27\pm0.09) \, \rm{GeV}$ is in excellent agreement with the experimental data $4273.3 \pm 8.3 ^{+17.2}_{-3.6} \mbox{ MeV}$  from the LHCb collaboration \cite{LHCb-16061,LHCb-16062}, which supports assigning the $X(4274)$ to be the $[sc]_A[\bar{s}\bar{c}]_V-[sc]_V[\bar{s}\bar{c}]_A$ type axialvector tetraquark state $X^+$ with a relative P-wave between the diquark and antidiquark constituents.

In the non-relativistic quark models,   naively we expect that the wave functions of the P-wave excitations vanish at the origin. In the present case, the pole residues have the relation $\lambda_{X^+}\ll\lambda_{X^-}$, the effect of the P-wave between the diquark and antidiquark constituents manifests itself, which is consistent with our naive expectation.

From Eq.\eqref{mass}, we can see that the masses have the relation  $M_{X^+}< M_{X^-}$. If we use the $\vec{S}_A$ and $\vec{S}_V$ to represent the spins of the axialvector and vector  diquarks (or antidiquarks) respectively,  the effective Hamiltonian contains  a term $\frac{1}{2}b\vec{L}\cdot\vec{L}+2a\vec{L}\cdot\vec{S}$, where $\vec{S}=\vec{S}_A+\vec{S}_V$, the $\vec{L}$ is the relative angular momentum \cite{Maiani-LS}. In the case  $X^-$,   $L=0$ and $\frac{1}{2}b\vec{L}\cdot\vec{L}+2a\vec{L}\cdot\vec{S}=0$.
 In the case $X^+$, the total spin $\vec{J}=\vec{L}+\vec{S}$, $J=1$ and $L=1$, the term $\frac{1}{2}b\vec{L}\cdot\vec{L}+2a\vec{L} \cdot \vec{S}=b+a\left[J(J+1)-L(L+1)-S(S+1)\right]=b-aS(S+1)=b$, $b-2a$ and $b-6a$ for  $S=0$, $1$ and $2$, respectively. If the spin-orbit coupling is strong enough,  the  $b-2a$ and $b-6a$ can have negative values, the effect of the additional P-wave  leads  to smaller tetraquark mass. At the present time, we have rare experimental data to fit the parameters  $a$ and $b$ if the vector diquarks are involved,  the calculations based on the QCD sum rules indicate that $M_{X^+}< M_{X^-}$.

\section{The width of the $X(4274)$ as the  axialvector tetraquark state }

We can study the hadronic coupling constant $g_{X^+J/\psi \phi}$  with the  three-point correlation function
$\Pi_{\alpha\beta\mu\nu}(p,q)$,
\begin{eqnarray}
\Pi_{\alpha\beta\mu\nu}(p,q)&=&i^2\int d^4xd^4y e^{ipx}e^{iqy}\langle 0|T\left\{J_\alpha^{J/\psi}(x)J_{\beta}^{\phi}(y)J_{\mu\nu}^{\dagger}(0)\right\}|0\rangle\, ,
\end{eqnarray}
where the currents
\begin{eqnarray}
J_\alpha^{J/\psi}(x)&=&\bar{c}(x)\gamma_\alpha c(x) \, ,\nonumber \\
J_\beta^{\phi}(y)&=&\bar{s}(y)\gamma_\beta s(y) \, ,
\end{eqnarray}
interpolate the mesons $J/\psi$ and $\phi(1020)$  respectively,
\begin{eqnarray}
\langle0|J_{\alpha}^{J/\psi}(0)|J/\psi(p)\rangle&=&f_{J/\psi}m_{J/\psi}\xi_\alpha \,\, , \nonumber \\
\langle0|J_{\beta}^{\phi}(0)|\phi(q)\rangle&=&f_{\phi}m_{\phi}\zeta_\beta \,\, ,
\end{eqnarray}
the $f_{J/\psi}$  and $f_{\phi}$ are the decay constants, the $\xi_\alpha$  and $\zeta_\beta$ are polarization vectors of the mesons $J/\psi$  and $\phi(1020)$, respectively.

At the phenomenological side,  we insert  a complete set of intermediate hadronic states with
the same quantum numbers as the current operators $J_\alpha^{J/\psi}(x)$, $J_\beta^{\phi}(y)$, $J_{\mu\nu}^{\dagger}(0)$ into the three-point
correlation function $\Pi_{\alpha\beta\mu\nu}(p,q)$ \cite{SVZ79,Reinders85}, and  isolate the ground state
contributions to obtain the  result,
\begin{eqnarray}
\Pi_{\alpha\beta\mu\nu}(p,q)&=&  \frac{f_{\phi}m_{\phi} f_{J/\psi}m_{J/\psi}\overline{\lambda}_{X^+}g_{X^+J/\psi \phi} }{(m_{X^+}^2-p^{\prime2})(m_{J/\psi}^2-p^2)(m_{\phi}^2-q^2)} \,\varepsilon^{\lambda\tau\rho\theta}p^{\prime}_\lambda\left(-g_{\alpha\rho}+\frac{p_{\alpha}p_{\rho}}{p^{ 2}} \right)\left(-g_{\beta\theta}+\frac{q_{\beta}q_{\theta}}{q^{ 2}} \right)\nonumber\\
&&\left[\left(-g_{\mu\tau}+\frac{p_{\mu}^{\prime}p^{\prime}_{\tau}}{p^{\prime 2}} \right) p^\prime_\nu-\left(-g_{\nu\tau}+\frac{p_{\mu}^{\prime}p^{\prime}_{\tau}}{p^{\prime 2}} \right) p^\prime_\mu\right]    \nonumber\\
&&+  \frac{f_{\phi}m_{\phi} f_{J/\psi}m_{J/\psi}\overline{\lambda}_{X^-}\langle J/\psi(p,\xi)\phi(q,\zeta)|X^-(p^\prime,\varepsilon)\rangle }{(m_{X^-}^2-p^{\prime2})(m_{J/\psi}^2-p^2)(m_{\phi}^2-q^2)} \,\xi_\alpha\zeta_\beta\,\varepsilon_{\mu\nu\rho\sigma}\,\varepsilon^{*\rho}p^{\prime\sigma}+\cdots  \nonumber\\
&=&\left\{ \frac{f_{\phi}m_{\phi} f_{J/\psi}m_{J/\psi}\overline{\lambda}_{X^+}g_{X^+J/\psi \phi} }{(m_{X^+}^2-p^{\prime2})(m_{J/\psi}^2-p^2)(m_{\phi}^2-q^2)} + \frac{1}{(m_{X^+}^2-p^{\prime2})(m_{J/\psi}^2-p^2)} \int_{s^0_\phi}^\infty dt\frac{\rho_{X^+\phi^\prime}(p^{\prime 2},p^2,t)}{t-q^2}\right.\nonumber\\
&&\left.+ \frac{1}{(m_{X^+}^2-p^{\prime2})(m_{\phi}^2-q^2)} \int_{s^0_{J/\psi}}^\infty dt\frac{\rho_{X^+\psi^\prime}(p^{\prime 2},t,q^2)}{t-p^2}\right. \nonumber\\
&&\left.+ \frac{1}{(m_{J/\psi}^2-p^{2})(m_{\phi}^2-q^2)} \int_{s^0_{X}}^\infty dt\frac{\rho_{X^{+\prime}J/\psi}(t,p^2,q^2)+\rho_{X^{+\prime}\phi}(t,p^2,q^2)}{t-p^{\prime2}}+\cdots\right\}\nonumber\\
&&\left(\varepsilon_{\alpha\beta\mu\lambda}p^{\lambda}p_\nu-\varepsilon_{\alpha\beta\nu\lambda}p^{\lambda}p_\mu+\cdots\right) +\cdots\nonumber\\
&=&\Pi(p^{\prime2},p^2,q^2) \, \left(\varepsilon_{\alpha\beta\mu\lambda}p^{\lambda}p_\nu-\varepsilon_{\alpha\beta\nu\lambda}p^{\lambda}p_\mu\right)+\cdots\, ,
\end{eqnarray}
where $p^\prime=p+q$, $\overline{\lambda}_{X^\pm}=\frac{\lambda_{X^\pm}}{m_{X^\pm}}$, $m_{X^+}=M_{X^+}$ the $g_{X^+J/\psi\phi}$  is the hadronic coupling constant defined by
\begin{eqnarray}
\langle J/\psi(p,\xi)\phi(q,\zeta)|X^+(p^{\prime},\varepsilon)\rangle&=&i g_{X^+J/\psi\phi} \, \varepsilon^{\lambda\tau\rho\theta} p^\prime_\lambda \varepsilon_\tau \xi^*_\rho \zeta^*_\theta\, ,
\end{eqnarray}
the four   functions $\rho_{X^+\phi^\prime}(p^{\prime 2},p^2,t)$, $ \rho_{X^+\psi^\prime}(p^{\prime 2},t,q^2)$,
$ \rho_{X^{+\prime}J/\psi}(t^\prime,p^2,q^2)$ and $\rho_{X^{+\prime}\phi}(t^\prime,p^2,q^2)$
   have complex dependence on the transitions between the ground states and the higher resonances  or the continuum states.

In this article, we choose the tensor structure $\varepsilon_{\alpha\beta\mu\lambda}p^{\lambda}p_\nu-\varepsilon_{\alpha\beta\nu\lambda}p^{\lambda}p_\mu$  to study the  hadronic coupling constant $g_{X^+J/\psi\phi}$   to avoid the contamination from the vector tetraquark state $X^-$, as the  tetraquark state $X^-$ is associated with the tensor structure $\varepsilon_{\mu\nu\bullet\bullet }$, where the  $\bullet\bullet$ denotes some functions of the $p$, $p^\prime$, $q$.
Furthermore, the contaminations originate from the   scalar meson $\chi_{c0}(3414)$ and scalar meson $f_0(980)$  are also avoided,
\begin{eqnarray}
\langle0|J_{\alpha}^{J/\psi}(0)|\chi_{c0}(p)\rangle&=&f_{\chi_{c0}}p_\alpha \,\, , \nonumber \\
\langle0|J_{\beta}^{\phi}(0)|f_0(q)\rangle&=&f_{f_0}q_\beta \,\, ,
\end{eqnarray}
 where the $f_{\chi_{c0}}$ and $f_{f_0}$  are the decay constants of the $\chi_{c0}(3414)$ and $f_0(980)$, respectively. Thereafter we will smear the superscript $+$ in the $X^+$ for simplicity.

The correlation function $\Pi(p^{\prime2},p^2,q^2)$ at the phenomenological side can be written as
\begin{eqnarray}
\Pi_H(p^{\prime2},p^2,q^2)&=&\int_{(m_{J/\psi}+m_{\phi})^2}^{s_{X}^0}ds^\prime \int_{4m_c^2}^{s^0_{J/\psi}}ds \int_0^{u^0_{\phi}}du  \frac{\rho_H(s^\prime,s,u)}{(s^\prime-p^{\prime2})(s-p^2)(u-q^2)}+\cdots\, ,
\end{eqnarray}
 through the dispersion relation, where the $\rho_H(s^\prime,s,u)$   is the hadronic spectral density,
\begin{eqnarray}
\rho_H(s^\prime,s,u)&=&{\lim_{\epsilon_3\to 0}}\,\,{\lim_{\epsilon_2\to 0}} \,\,{\lim_{\epsilon_1\to 0}}\,\,\frac{ {\rm Im}_{s^\prime}\, {\rm Im}_{s}\,{\rm Im}_{u}\,\Pi_H(s^\prime+i\epsilon_3,s+i\epsilon_2,u+i\epsilon_1) }{\pi^3} \, ,
\end{eqnarray}
we introduce the subscript $H$ to denote the hadron side.

We carry out  the operator product expansion for the correlation function $\Pi_{\alpha\beta\mu\nu}(p,q)$ up to the vacuum condensates of dimension 5 and neglect the tiny  contributions of the gluon condensate.  We contract the quark fields $s$ and $c$ in the correlation function
$\Pi_{\alpha\beta\mu\nu}(p,q)$ with Wick theorem, and obtain the result,
\begin{eqnarray}
\Pi_{\alpha\beta\mu\nu}(p,q)&=&\frac{\varepsilon^{ijk}\varepsilon^{imn}}{\sqrt{2}} \int d^4x d^4y e^{ip\cdot x}e^{iq\cdot y} \nonumber\\
&&\Big\{{\rm Tr} \Big[\gamma_\alpha S^{ak}_c(x)\gamma_5\gamma_\nu CS^{Tbj}(y)C\gamma_\beta CS^{Tmb}(-y)C\gamma_\mu S_c^{na}(-x) \Big]    \nonumber\\
&& +{\rm Tr} \Big[\gamma_\alpha S^{ak}_c(x)\gamma_\mu CS^{Tbj}(y)C\gamma_\beta CS^{Tmb}(-y)C\gamma_\nu\gamma_5 S_c^{na}(-x) \Big]  \Big\} \, ,
\end{eqnarray}
where the $a$, $b$, $i$, $j$, $k$, $m$ and $n$ are color indexes, the $S_c^{ak}(x)$ and $S^{mb}(x)$ are the full $c$ and $s$ quark propagators, respectively, see Eqs.\eqref{Propagator-L}-\eqref{Propagator-H}.  Then we compute  the integrals both in the coordinate space and in the momentum space,  and obtain the correlation function at the QCD side, therefore the QCD spectral density through dispersion relation,
\begin{eqnarray}
\Pi_{QCD}(p^{\prime 2},p^2,q^2)&=&  \int_{4m_c^2}^{s^0_{J/\psi}}ds \int_0^{u^0_{\phi}}du  \frac{\rho_{QCD}(p^{\prime2},s,u)}{(s-p^2)(u-q^2)}+\cdots\, ,
\end{eqnarray}
 where the $\rho_{QCD}(p^{\prime 2},s,u)$   is the QCD spectral density,
\begin{eqnarray}
\rho_{QCD}(p^{\prime 2},s,u)&=& {\lim_{\epsilon_2\to 0}} \,\,{\lim_{\epsilon_1\to 0}}\,\,\frac{  {\rm Im}_{s}\,{\rm Im}_{u}\,\Pi_{QCD}(p^{\prime 2},s+i\epsilon_2,u+i\epsilon_1) }{\pi^2} \, ,
\end{eqnarray}
we introduce the subscript $QCD$ to denote the $QCD$ side.
However,  the QCD spectral density $\rho_{QCD}(s^\prime,s,u)$ does  not exist,
\begin{eqnarray}
\rho_{QCD}(s^\prime,s,u)&=&{\lim_{\epsilon_3\to 0}}\,\,{\lim_{\epsilon_2\to 0}} \,\,{\lim_{\epsilon_1\to 0}}\,\,\frac{ {\rm Im}_{s^\prime}\, {\rm Im}_{s}\,{\rm Im}_{u}\,\Pi_{QCD}(s^\prime+i\epsilon_3,s+i\epsilon_2,u+i\epsilon_1) }{\pi^3} \nonumber\\
&=&0\, ,
\end{eqnarray}
because
\begin{eqnarray}
{\lim_{\epsilon_3\to 0}}\,\,\frac{ {\rm Im}_{s^\prime}\,\Pi_{QCD}(s^\prime+i\epsilon_3,p^2,q^2) }{\pi} &=&0\, .
\end{eqnarray}

We math the hadron side  with the QCD side of the correlation function,
and carry out the integral over $ds^\prime$  firstly to obtain the solid duality \cite{WangZhang-Solid},
\begin{eqnarray}
\int_{\Delta_s^2}^{s^0} ds \int_{\Delta_u^2}^{u^0} du \frac{\rho_{QCD}(p^{\prime2},s,u)}{(s-p^2)(u-q^2)}&=&\int_{\Delta_s^2}^{s^0} ds \int_{\Delta_u^2}^{u^0} du \frac{1}{(s-p^2)(u-q^2)}\left[ \int_{\Delta^2}^{\infty} ds^\prime \frac{\rho_{H}(s^{\prime},s,u)}{s^\prime-p^{\prime2}}\right]\, , \nonumber\\
\end{eqnarray}
 the $\Delta_s^2$ and $\Delta_u^2$ are the thresholds $4m_c^2$ and $0$ respectively, the $\Delta^2$ is the threshold $(m_{J/\psi}+m_{\phi})^2$.
 Now we write the quark-hadron duality explicitly,
 \begin{eqnarray}
  \int_{4m_c^2}^{s^0_{J/\psi}}ds \int_0^{u^0_{\phi}}du  \frac{\rho_{QCD}(p^{\prime2},s,u)}{(s-p^2)(u-q^2)}&=& \int_{4m_c^2}^{s^0_{J/\psi}}ds \int_0^{u^0_{\phi}}du  \int_{(m_{J/\psi}+m_{\phi})^2}^{\infty}ds^\prime \frac{\rho_H(s^\prime,s,u)}{(s^\prime-p^{\prime2})(s-p^2)(u-q^2)} \nonumber\\
  &=&\frac{f_{\phi}m_{\phi} f_{J/\psi}m_{J/\psi}\overline{\lambda}_{X}g_{XJ/\psi \phi} }{(m_{X}^2-p^{\prime2})(m_{J/\psi}^2-p^2)(m_{\phi}^2-q^2)}  +\frac{C_{X^{\prime}J/\psi}+C_{X^{\prime}\phi}}{(m_{J/\psi}^2-p^{2})(m_{\phi}^2-q^2)} \, ,\nonumber\\
\end{eqnarray}
we introduce the parameters  $C_{X^\prime\phi}$ and $C_{X^\prime J/\psi}$   to parameterize the net effects,
\begin{eqnarray}
C_{X^\prime\phi}&=&\int_{s^0_{X}}^\infty dt\frac{ \rho_{X^\prime\phi}(t,p^2,q^2)}{t-p^{\prime2}}\, ,\nonumber\\
C_{X^\prime J/\psi}&=&\int_{s^0_{X}}^\infty dt\frac{ \rho_{X^\prime J/\psi}(t,p^2,q^2)}{t-p^{\prime2}}\, .
\end{eqnarray}
 No approximation is needed, we do not need the continuum threshold parameter $s^0_{X}$ in the $s^\prime$ channel. The present approach was introduced in Ref.\cite{WangZhang-Solid}.

In numerical calculations,   we   take the unknown functions  $C_{X^\prime\phi}$ and $C_{X^\prime J/\psi}$  as free parameters, and choose the suitable values  to eliminate the contaminations from the higher resonances (i.e. $X^\prime$ et al) and continuum states to obtain the stable QCD sum rules with the variations of
the Borel parameters.
We  set $p^{\prime2}=p^2$  and perform the double Borel transform  with respect to the variables $P^2=-p^2$ and $Q^2=-q^2$ respectively  to obtain the  QCD sum rules,
\begin{eqnarray}
&& \frac{f_{\phi}m_{\phi} f_{J/\psi}m_{J/\psi}\overline{\lambda}_{X}g_{XJ/\psi \phi}}{m_{X}^2-m_{J/\psi}^2} \left[ \exp\left(-\frac{m_{J/\psi}^2}{T_1^2} \right)-\exp\left(-\frac{m_{X}^2}{T_1^2} \right)\right]\exp\left(-\frac{m_{\phi}^2}{T_2^2} \right) \nonumber\\
&&+\left(C_{X^{\prime}J/\psi}+C_{X^{\prime}\phi}\right) \exp\left(-\frac{m_{J/\psi}^2}{T_1^2} -\frac{m_{\phi}^2}{T_2^2} \right)\nonumber\\
&&=-\frac{1}{48\sqrt{2}\pi^4}\int_{4m_c^2}^{s^0_{J/\psi}} ds \int_{0}^{s^0_{\phi}} du  u\sqrt{1-\frac{4m_c^2}{s}}\left(1+\frac{2 m_c^2}{s} \right)\exp\left(-\frac{s}{T_1^2} -\frac{u}{T_2^2} \right)\nonumber\\
&&+\frac{m_s  \langle\bar{s}s\rangle}{6\sqrt{2}\pi^2} \int_{4m_c^2}^{s^0_{J/\psi}} ds \sqrt{1-\frac{4m_c^2}{s}}\left(1+\frac{2 m_c^2}{s} \right)\exp\left(-\frac{s}{T_1^2}  \right) \nonumber\\
&&-\frac{m_s \langle\bar{s}g_s\sigma Gs\rangle}{72\sqrt{2}\pi^2 T_2^2} \int_{4m_c^2}^{s^0_{J/\psi}} ds \sqrt{1-\frac{4m_c^2}{s}}\left(1+\frac{2 m_c^2}{s} \right)\exp\left(-\frac{s}{T_1^2}  \right)   \, .
\end{eqnarray}

The hadronic parameters are taken as    $m_{\phi}=1.019461\,\rm{GeV}$,
$m_{J/\psi}=3.0969\,\rm{GeV}$ \cite{PDG},
$f_{J/\psi}=0.418 \,\rm{GeV}$  \cite{Becirevic}, $f_{\phi}=0.253\,\rm{GeV}$, $\sqrt{s^0_{\phi}}=1.5\,\rm{GeV}$ \cite{Wang-2016-Y4274}, $\sqrt{s^0_{J/\psi}}=3.6\,\rm{GeV}$,
$M_X=4273.3\,\rm{MeV}$ \cite{LHCb-16061,LHCb-16062},   $\lambda_{X}=1.52\times 10^{-2}\,\rm{GeV}^5$. At the QCD side, we can take the energy scale of the QCD spectral density to be $\mu=1\,\rm{GeV}$, just like in the two-point QCD sum rules. However, at the energy scale $\mu=1.0\,\rm{GeV}$, $2m_c(1\,\rm{GeV})=2.8\,\rm{GeV}$, $\sqrt{s^0_{J/\psi}}-2m_c(1\,\rm{GeV})=0.8\,\rm{GeV}$, the integral interval $4m_c^2-s^0_{J/\psi}$ is too small to obtain stable QCD sum rules; the interval $\sqrt{s^0_{J/\psi}}-2m_c(\mu)$ should be larger than $1\,\rm{GeV}$ to obtain stable
QCD sum rules. At the energy scale $\mu=m_c(m_c)=1.275\pm0.025\,\rm{GeV}$, $\sqrt{s^0_{J/\psi}}-2m_c(m_c)=1.05\pm0.05\,\rm{GeV}$, the lower bound is $1.0\,\rm{GeV}$, the uncertainty is out of control.  So in this article, we choose the typical energy scale
$\mu=m_c(m_c)$ and neglect the uncertainties of the quark masses. It is the shortcoming of the present QCD sum rules, we can only obtain  qualitative conclusion,
 as rigorous uncertainty  analysis is lack. We set the Borel parameters to be $T_1^2=T_2^2=T^2$ for simplicity.
The unknown parameters are chosen as $C_{X^{\prime}J/\psi}+C_{X^{\prime}\phi}=-0.00145\,\rm{GeV}^6 $   to obtain  platform in the Borel window $T^2=(2.8-3.8)\,\rm{GeV}^2$.
In calculations, we observe that the predicted hadronic coupling constant  $g_{XJ/\psi \phi}$ increases  monotonously  with increase of the energy scale. The energy scale $\mu=m_c(m_c)=1.275\,\rm{GeV}$ is an acceptable energy scale in the QCD sum rules for the $J/\psi$ and $\phi(1020)$, although it deviates slightly from  the optimal energy scale $\mu=1\,\rm{GeV}$ in the QCD sum rules for the $X(4274)$; the deviation leads to unavoidable uncertainty in the hadronic coupling constant  $g_{XJ/\psi \phi}$, i.e. we underestimate the value of the  $g_{XJ/\psi \phi}$ slightly.

In Fig.3, we plot the hadronic coupling constant  $g_{XJ/\psi \phi}$  with variation of the  Borel parameter $T^2$. From the figure, we can see that there appears platform in the Borel window indeed, where the uncertainty originates from the Borel parameter $T^2$ is small and can be neglected safely.
The central value of the  hadronic coupling constant   $g_{XJ/\psi \phi}$,
\begin{eqnarray}
g_{XJ/\psi \phi} &=&-1.05\, ,
\end{eqnarray}
which corresponds  to the central values of all the input parameters. We obtain the
 decay width,
\begin{eqnarray}
\Gamma(X(4274)\to J/\psi \phi)&=& \frac{p\left(m_{X},m_{J/\psi},m_{\phi}\right)}{24\pi m_{X}^2}g_{XJ/\psi\phi}^2\left\{\frac{\left(m_{X}^2-m_{\phi}^2\right)^2}{2m_{J/\psi}^2}+\frac{\left(m_{X}^2-m_{J/\psi}^2\right)^2}{2m_{\phi}^2} \right.\nonumber\\
&&\left. +4m_X^2-\frac{m_{J/\psi}^2+m_\phi^2}{2}\right\} \nonumber\\
&=&47.9\,{\rm{MeV}}\sim 56 \pm 11 ^{+8}_{-11} {\mbox{ MeV}}\,\,\,\rm{Experimental\,\,\,value} \,\,\cite{LHCb-16061,LHCb-16062} \, ,
\end{eqnarray}
where $p(a,b,c)=\frac{\sqrt{[a^2-(b+c)^2][a^2-(b-c)^2]}}{2a}$.	The width $\Gamma(X(4274)\to J/\psi \phi)=47.9\,{\rm{MeV}}$ is in excellent agreement with the experimental data $56 \pm 11 ^{+8}_{-11} {\mbox{ MeV}}$ from the LHCb collaboration \cite{LHCb-16061,LHCb-16062}.
The present work supports assigning the $X(4274)$ to be the  $[sc]_A[\bar{s}\bar{c}]_V-[sc]_V[\bar{s}\bar{c}]_A$  type tetraquark state with a relative P-wave between the diquark and antidiquark constituents.

\begin{figure}
 \centering
 \includegraphics[totalheight=7cm,width=10cm]{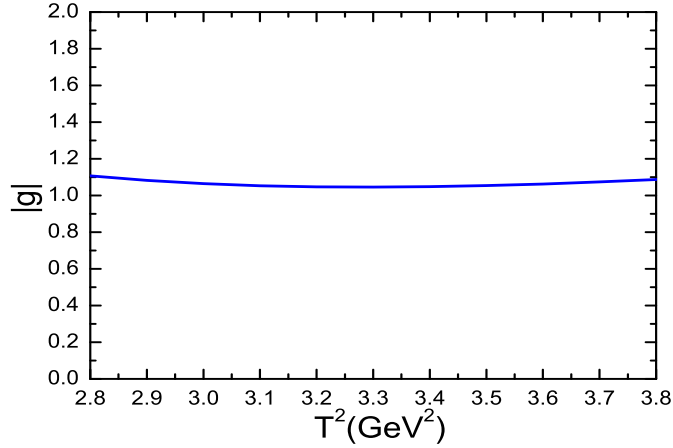}
   \caption{ The hadronic coupling constant $g_{XJ/\psi\phi}$ with variation of the Borel parameter $T^2$.   }
\end{figure}

In Ref.\cite{Wang-2016-Y4274}, we  construct the  color octet-octet type axialvector current $\eta_\mu(x)$ to study the mass and width of the  $X(4274)$, \begin{eqnarray}
\eta_\mu(x) &=&\frac{1}{\sqrt{2}}\Big[\bar{s}(x)i\gamma_5\lambda^ac(x)\,\bar{c}(x)\gamma_\mu\lambda^a s(x)-\bar{s}(x)\gamma_\mu \lambda^ac(x)\,\bar{c}(x)i\gamma_5 \lambda^as(x) \Big] \, .
\end{eqnarray}
Now we perform Fierz re-arrangement both in the color and Dirac-spinor  spaces and obtain the  result,
\begin{eqnarray}
\eta_\mu&=&-\frac{N_c+1}{N_c}\Big\{\frac{i}{2\sqrt{2}} \left(s^TC\gamma_5t^Ac\,\bar{s}\gamma_\mu Ct^A\bar{c}^T+s^TC\gamma_\mu t^Ac\,\bar{s}\gamma_5 Ct^A\bar{c}^T\right)\nonumber\\
&&+\frac{1}{2\sqrt{2}} \left(s^TC\gamma^\alpha\gamma_5t^Ac\,\bar{s}\sigma_{\alpha\mu} Ct^A\bar{c}^T-s^TC\sigma_{\alpha\mu} t^Ac\,\bar{s}\gamma_5 \gamma^\alpha Ct^A\bar{c}^T\right)\Big\}\nonumber\\
&&+\frac{N_c-1}{N_c}\Big\{\frac{i}{2\sqrt{2}} \left(s^TC\gamma_5t^Sc\,\bar{s}\gamma_\mu Ct^S\bar{c}^T+s^TC\gamma_\mu t^Sc\,\bar{s}\gamma_5 Ct^S\bar{c}^T\right)\nonumber\\
&&+\frac{1}{2\sqrt{2}} \left(s^TC\gamma^\alpha\gamma_5t^Sc\,\bar{s}\sigma_{\alpha\mu} Ct^S\bar{c}^T-s^TC\sigma_{\alpha\mu} t^Sc\,\bar{s}\gamma_5 \gamma^\alpha Ct^S\bar{c}^T\right)\Big\} \nonumber\\
&=&-\frac{N_c+1}{N_c}\Big\{\frac{i}{2} J^1_\mu  +\frac{1}{2} J^4_\mu \Big\}+\frac{N_c-1}{N_c}\Big\{\frac{i}{2}\hat{J}^1_\mu  +\frac{1}{2} \hat{J}^4_\mu \Big\} \, ,
\end{eqnarray}
where
\begin{eqnarray}
\hat{J}^1_\mu &=&\frac{1}{\sqrt{2}} \left(s^TC\gamma_5t^Sc\,\bar{s}\gamma_\mu Ct^S\bar{c}^T+s^TC\gamma_\mu t^Sc\,\bar{s}\gamma_5 Ct^S\bar{c}^T\right)\, ,\nonumber\\
\hat{J}^4_\mu &=&\frac{1}{\sqrt{2}} \left(s^TC\gamma^\alpha\gamma_5t^Sc\,\bar{s}\sigma_{\alpha\mu} Ct^S\bar{c}^T-s^TC\sigma_{\alpha\mu} t^Sc\,\bar{s}\gamma_5 \gamma^\alpha Ct^S\bar{c}^T\right)\, .
\end{eqnarray}
The current $J^1_\mu(x)$ couples potentially to the $[sc]_S[\bar{s}\bar{c}]_A+[sc]_A[\bar{s}\bar{c}]_S$ type axialvector tetraquark state with a mass  $3.95\pm0.09\,\rm{GeV}$ \cite{Wang-2016-Y4140}, the current $J^4_\mu(x)$ couples potentially to the $[sc]_T[\bar{s}\bar{c}]_V-[sc]_V[\bar{s}\bar{c}]_T$  type axialvector tetraquark state with a mass $4.14\pm0.10 \,\rm{GeV}$ \cite{WangDi-1811}. While the currents $\hat{J}^1_\mu(x)$ and $\hat{J}^4_\mu(x)$ couple potentially to the $[sc]_S^6[\bar{s}\bar{c}]^{\bar{6}}_A+[sc]_A^6[\bar{s}\bar{c}]_S^{\bar{6}}$ type and $[sc]_T^6[\bar{s}\bar{c}]_V^{\bar{6}}-[sc]_V^6[\bar{s}\bar{c}]_T^{\bar{6}}$  type axialvector tetraquark states, respectively. The current $\eta_\mu(x)$ is a special superposition of the currents $J^1_\mu(x)$, $J^4_\mu(x)$, $\hat{J}^1_\mu(x)$ and $\hat{J}^4_\mu(x)$, and embodies the net effects. The ideal energy scales of the QCD spectral densities of the correlation functions for the currents $J^1_\mu(x)$ and $J^4_\mu(x)$ are
$\mu=1.5\,\rm{GeV}$ and $2.0\,\rm{GeV}$, respectively \cite{Wang-2016-Y4140,WangDi-1811}, while the ideal energy scale of the QCD spectral density of the correlation function for the current $\eta_\mu(x)$ is $\mu=1.45\,\rm{GeV}$ \cite{Wang-2016-Y4274}. The energy scale for the lowest tetraquark state is consistent with that for the color octet-octet type tetraquark molecule-like state, although the two energy scales are determined by very different $c$-quark mass ${\mathbb{M}}_c$.
There does not exist a $[sc]_A[\bar{s}\bar{c}]_V-[sc]_V[\bar{s}\bar{c}]_A$ type component in the current $\eta_\mu(x)$, the current $J_{\mu\nu}(x)$ chosen in the present work differs  from the current chosen in Ref.\cite{Wang-2016-Y4274} completely. Furthermore, the $[sc]_A[\bar{s}\bar{c}]_V-[sc]_V[\bar{s}\bar{c}]_A$ type and $[\bar{s}\lambda^ac]_P[\bar{c}\lambda^a s]-[\bar{s} \lambda^ac]_V[\bar{c}\lambda^as]$ type axialvector four-quark states have completely  different widths, which originate from the completely different quark structures.

\section{Conclusion}
In this article, we construct  the $[sc]_A[\bar{s}\bar{c}]_V-[sc]_V[\bar{s}\bar{c}]_A$ type tensor  current to study  the $X(4274)$ with the QCD sum rules by carrying out the operator product expansion up to the vacuum condensates of dimension 10. The tensor current couples potentially to both the $J^{PC}=1^{++}$ and $1^{-+}$ tetraquark states, we separate those contributions unambiguously by introducing  suitable projectors. In calculations, we use the energy scale formula to determine the optimal energy scales of the QCD spectral densities, and extract the masses of the $J^{PC}=1^{++}$ and $1^{-+}$ tetraquark states at different energy scales.       The predicted mass $M_{X}=(4.27\pm0.09) \, \rm{GeV}$ for the $J^{PC}=1^{++}$ tetraquark state  is in excellent agreement with the  experimental value $4273.3 \pm 8.3 ^{+17.2}_{-3.6} \mbox{ MeV}$  from the LHCb collaboration.
Then we study the two-body strong decay $X(4274)\to J/\psi\phi$ with the QCD sum rules based on the solid quark-hadron duality introduced in our previous work.
 The central value of the predicted width $\Gamma(X(4274)\to J/\psi \phi)=47.9\,{\rm{MeV}}$ is in excellent agreement with  the experimental value $56 \pm 11 ^{+8}_{-11} {\mbox{ MeV}}$ from the LHCb collaboration. In summary, the present work supports assigning the $X(4274)$ to be the $J^{PC}=1^{++}$  $[sc]_A[\bar{s}\bar{c}]_V-[sc]_V[\bar{s}\bar{c}]_A$  tetraquark state with a relative P-wave between the diquark and antidiquark constituents. Furthermore, we obtain the mass of the $[sc]_A[\bar{s}\bar{c}]_V-[sc]_V[\bar{s}\bar{c}]_A$ type tetraquark state with $J^{PC}=1^{-+}$ as a byproduct.

\section*{Acknowledgements}
This  work is supported by National Natural Science Foundation, Grant Number  11775079.

\end{document}